\documentclass[a4paper,11pt]{article}
\pdfoutput=1 
\usepackage{jheppub} 
\usepackage[utf8]{inputenc}
\usepackage[T1]{fontenc} 
\usepackage{bbold}
\usepackage{tabls}
\usepackage{multirow}
\usepackage{pifont}
\usepackage{cancel}
\usepackage{caption}
\usepackage{graphicx}
\usepackage{subcaption}
\usepackage{xcolor}

\newcommand{\bea}{\begin{eqnarray}}
\newcommand{\beq}{\begin{equation}}
\newcommand{\eea}{\end{eqnarray}}
\newcommand{\eeq}{\end{equation}}

\newcommand{\nn}{\nonumber}
\newcommand{\s}{\small}
\newcommand{\1}{{\bf 1}}
\newcommand{\3}{{\bf 3}}
\newcommand{\tr}{{\bf 3'}}
\newcommand{\4}{{\bf 4}}
\newcommand{\5}{{\bf 5}}
\newcommand{\ir}{{\bf r}}
\newcommand{\Oth}{$\theta~$}
\newcommand{\upmns}{$U_{\rm PMNS}~$}

\newcommand{\marisa}[1]{\textcolor{blue}{#1}}

\title{\boldmath Lepton Flavor Violation and Neutrino masses from $A_5$ and CP in the Non-Universal MSSM}
    
\author[a]{M.L. L\'opez-Ib\'a\~nez}
\author[b]{,\,Aurora Melis}
\author[a]{,\,Davide Meloni}
\author[b]{and Oscar Vives}

\affiliation[a]{Dipartimento di Matematica e Fisica, Università di Roma Tre and INFN, Sezione Roma Tre, \\
Via della Vasca Navale 84, 00146, Roma (Italy)}   
\affiliation[b]{Departament de F\'{i}sica T\`{e}orica, Universitat de Val\`{e}ncia and IFIC, Universitat de Val\`{e}ncia-CSIC,  \\
Dr. Moliner 50, E-46100 Burjassot (Val\`{e}ncia), Spain}

\emailAdd{maloi2@uv.es}
\emailAdd{aurora.melis@ific.uv.es}
\emailAdd{davide.meloni@uniroma3.it}
\emailAdd{oscar.vives@uv.es}

\preprint{FTUV-19-0114, IFIC-19-04}

\abstract{
We analyze the phenomenological consequences of embedding a flavor symmetry based on the groups $A_5$ and CP in a supersymmetric framework.
We concentrate on the leptonic sector, where two different residual symmetries are assumed to be conserved at leading order for charged and neutral leptons.
All possible realizations to generate neutrino masses at tree level are investigated.
Sizable flavor violating effects in the charged lepton sector are unavoidable due to the non-universality of soft-breaking terms determined by the symmetry.
We derive testable predictions for the neutrino spectrum, lepton mixing and flavor changing processes with non-trivial relations among observables.}

\begin{document}
\maketitle  
\setlength{\parindent}{0in}

\section{Introduction}
\label{sec:intoduction}

In recent years, different experiments have accumulated a wealth of experimental data on neutrino parameters that have allowed us to extract with reasonable precision the Pontecorvo-Maki-Nakagawa-Sakata (PMNS) mixing matrix \cite{Pontecorvo:1957cp, Pontecorvo:1957qd, Maki:1962mu, Pontecorvo:1967fh} and the neutrino mass differences.
Still, the determination of the absolute neutrino mass scale and the Dirac CP phase remain to be completed.
Future experiments like \texttt{DUNE} \cite{Acciarri:2015uup, Acciarri:2016ooe, Acciarri:2016crz, Strait:2016mof}, \texttt{T2HK} \cite{Abe:2015zbg}, \texttt{T2HKK} \cite{Abe:2016ero} and \texttt{NO$\nu$A} \cite{NOvACP} will shed light on these quantities; however, even a full determination of the neutrino mass matrix $m_\nu$ will not be enough to fix the mechanism responsible for it and uncover the origin of the observed flavor patterns.
Although this is also true in the case of the quark and charged lepton Yukawa couplings, the smallness of neutrino masses and its favorite explanation through a seesaw mechanism makes this problem specially critical. \\
If neutrinos are Majorana particles, their masses are well described through a $d=5$ Weinberg operator \cite{PhysRevLett.43.1566} in the Standard Model (SM).
But there exist different possibilities to generate this effective operator from a more fundamental theory at higher energies, like type I, type II or type III seesaw, radiative mass models, etc.
It is clear that the measurement of neutrino masses and mixing angles alone will not be enough to discriminate among these alternative mechanisms and to infer the couplings responsible for them.
For instance, for a type-I seesaw mechanism, both the neutrino Yukawa couplings and the right-handed neutrino Majorana mass would combine to generate the Weinberg operator but the SM does not provide information to disentangle them from the available experimental data.\\
Nevertheless, if flavor dependent new physics is close to the electroweak scale, as naturally expected in most of the extensions of the SM, it will provide additional information on flavor dynamics helping us to inspect the mechanism responsible for neutrino masses and to determine the parameters of the model.
One example of this, which has been explored in previous works \cite{Tsumura:2009yf, Buras:2011wi, Calibbi:2012at, Koide:2012uf, Bao:2015pva, Pascoli:2016wlt, Heinrich:2018nip} but we do not consider here, supposes that the flavor symmetry is broken around the electroweak scale.
In that case, the scalar flavons may mediate lepton flavor violating (LFV) processes in a measurable way while the fields themselves could be produced and detected in future colliders.
Another possibility is Supersymmetry (SUSY), which we consider the perfect example for this as it generically contains new flavor interactions in its soft-breaking sector in the presence of a flavor symmetry.
As shown in \cite{Das:2016czs, Lopez-Ibanez:2017xxw, deMedeirosVarzielas:2018vab, Calibbi:2012yj, Antusch:2011sq}, non-trivial flavor structures are unavoidable if the scale of transmission of SUSY breaking to the visible sector, $\Lambda_{\rm SUSY}$, is larger than the scale of breaking of the flavor symmetry, $\Lambda_f$.\\
The group $A_5$ combined with the so-called generalized CP symmetry has already been studied in different contexts \cite{Li:2015jxa, DiIura:2015kfa, Ballett:2015wia, Turner:2015uta, DiIura:2018fnk}.
Here, we aim to extend the work in \cite{DiIura:2018fnk} and analyze the phenomenological implications of considering $A_5$ and CP in a supersymmetric model.
In these conditions, we will show that constraints from lepton flavor violation are very strong and, in many cases, they are able to explore supersymmetric masses well beyond the reach of direct searches at \texttt{LHC} \cite{SUSYCMS, SUSYATLAS}. Besides, if SUSY is found in future experiments, we will obtain additional information on the structure of flavor matrices that will help us to distinguish between the different mechanisms responsible for neutrino masses.\\
The paper is organized as follows: in Section \ref{sec:reviewSUSY} and \ref{sec:reviewA5CP}, we revisit the implications of introducing a flavor symmetry in SUSY and the main features of $A_5$ and CP as a flavor group; Section \ref{sec:KahlerSoft} is dedicated to derive the minimal set of flavor-conserving operators entering the Kähler potential and the soft-mass terms; in Section \ref{sec:results}, the phenomenology of the model is analyzed; finally, we conclude in Section \ref{sec:conclusions} summarizing the most important results.

\section{\boldmath Flavor symmetries in supersymmetric theories}
\label{sec:reviewSUSY}

\begin{figure}[t!]
    \begin{subfigure}[h]{\textwidth}
    \centering
        \includegraphics[width=0.45\textwidth]{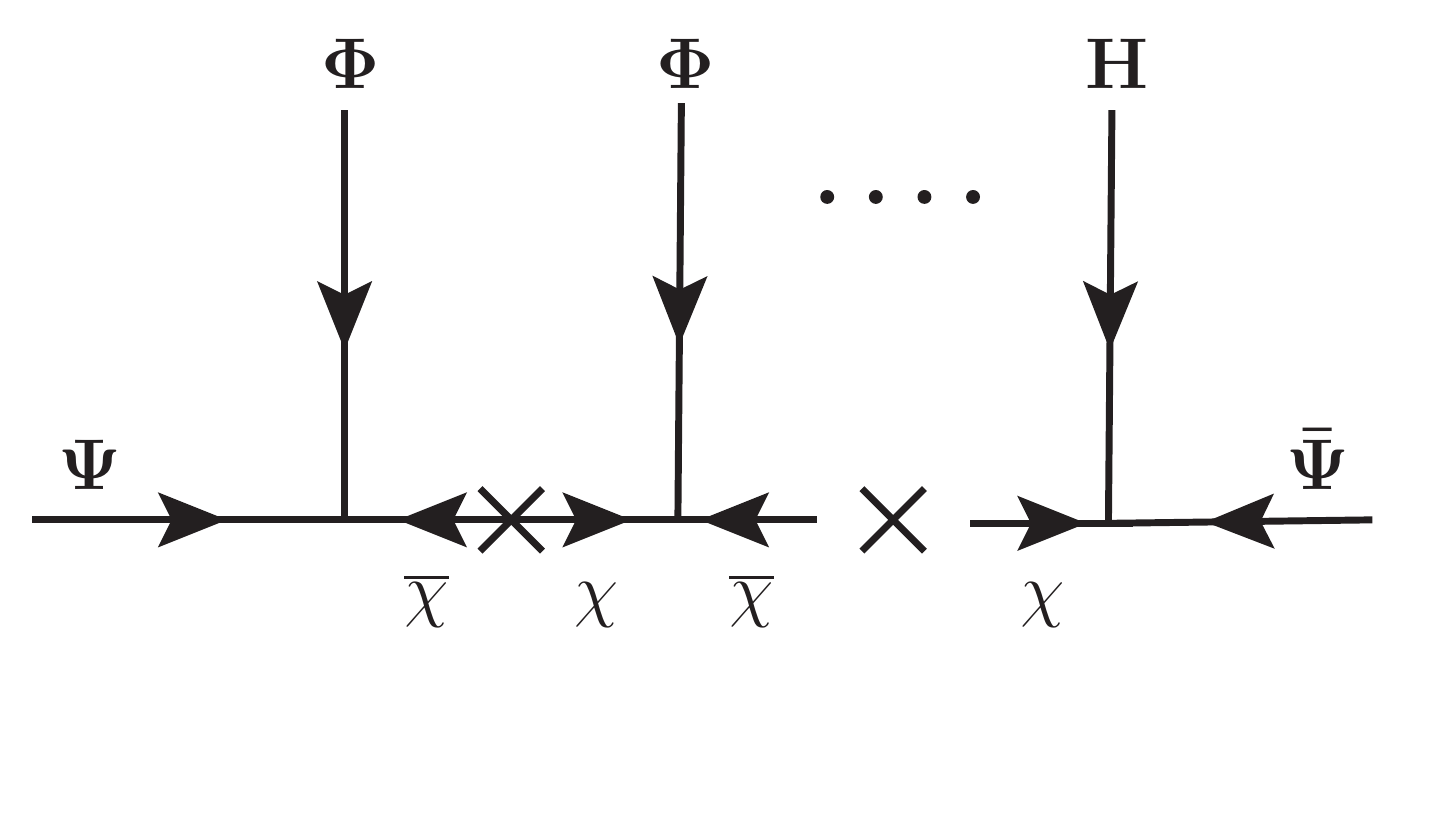}
        \subcaption{}
    \end{subfigure}
    \begin{subfigure}[h]{\textwidth}
    \centering
        \includegraphics[width=0.55\textwidth]{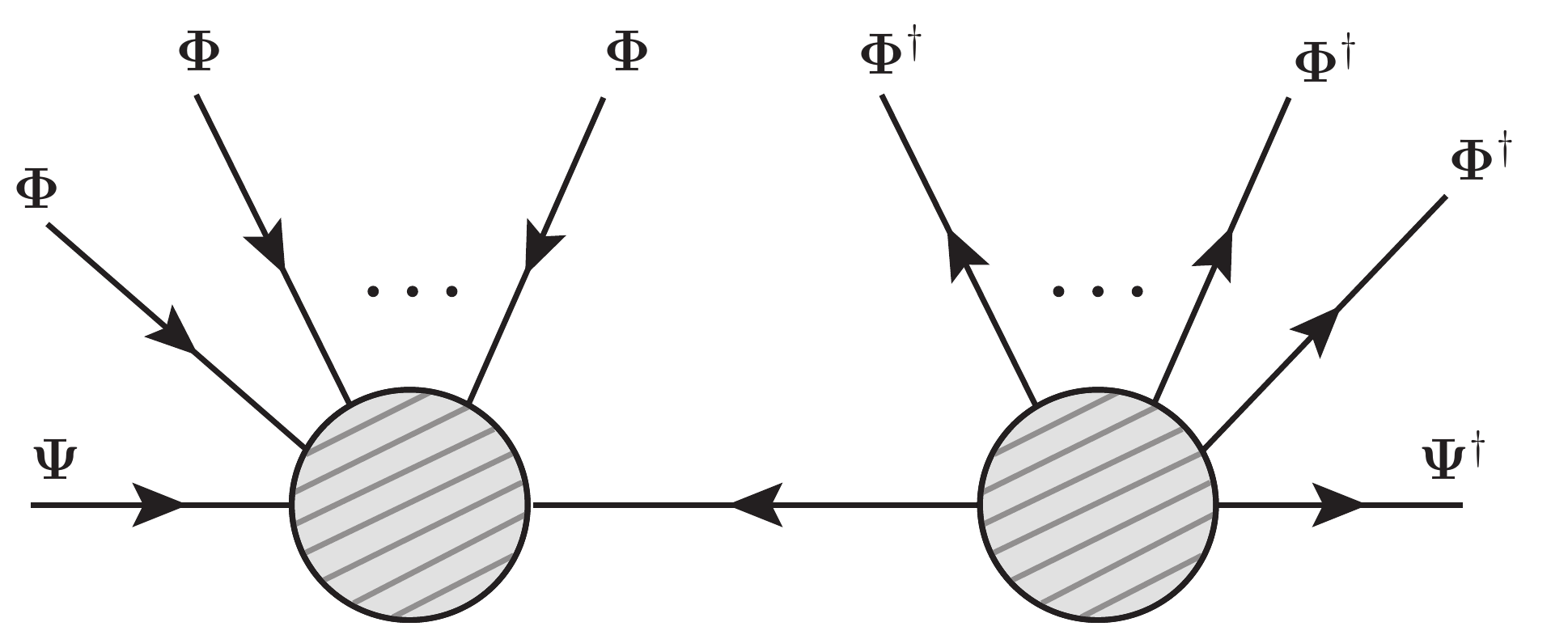}
        \subcaption{}
    \end{subfigure}
    \caption{\label{fig:W&K_flavor} \small 
    (a) A supergraph representation of the processes that generate the corrections to the Superpotential in the second term of Eq.\eqref{eqn:superpot}.
    It involves $n_{\rm in}$ flavon insertions.
    Internal lines represent the heavy mediators $\chi$ while the crosses stand for SUSY mass insertions $M_\chi$.\\
    (b) A supergraph depiction of the processes that generate the correction to the Kähler potential in the second term of Eq.\eqref{eqn:kahler}.
    Each bubble with entering (leaving) lines represent a set of $n_{\rm in}$ ($n_{\rm out}$) fields (dagger fields) in the same fashion than the upper diagram, mediated by $\chi$ heavy superfields.}
\end{figure}

The embedding of a flavor symmetry in a supersymmetric theory implies that the different superfields have definite transformation properties under the flavor symmetry and, then, the whole Lagrangian in terms of component fields is necessarily invariant under this symmetry.
Initially, the SM Yukawa couplings are forbidden and they are only generated after spontaneous breaking of the symmetry \cite{Froggatt:1978nt}.
Similarly, the flavor structures of the soft-breaking terms will be determined in terms of the flavon vevs \cite{Das:2016czs, Lopez-Ibanez:2017xxw, deMedeirosVarzielas:2018vab, Calibbi:2012yj, Antusch:2011sq}.\\
Sizable non-universal contributions to the soft-terms appear in the low-energy effective theory if the scale of mediation of SUSY breaking, $\Lambda_{\rm SUSY}$, is above the scale of flavor symmetry beaking $\Lambda_{f}$, $\Lambda_{\rm SUSY} \gg \Lambda_{f}$.
Supergravity serves as illustrative example to show this, although the results outlined here are more general \cite{Das:2016czs}.
In the following, we consider mSUGRA, which depends only on five input parameters and gives rise to the Minimal Supersymmetric Standard Model (MSSM) at low energies.\\
In Supergravity, SUSY breaking is propagated to the visible sector through gravitational interactions, suppressed by the Planck scale $M_{\rm Pl}$.
mSUGRA is the simplest and most conservative scenario that parametrizes the breaking of supersymmetry by a single field, universally coupled to the visible sector, with a non-vanishing F-term, $\langle {\sf X}\rangle = F_{\sf X} \neq 0$.
In the full theory, before the breaking of the flavor symmetry, the soft-breaking operators are generated from those associated with the Yukawa couplings in the superpotential (${\cal W}_\Psi$) and kinetic terms in the Kähler potential ($K_\Psi$) through the insertion of the spurion field as
\begin{equation} \label{eqn:Lsoft}
    {\cal L}_{\rm soft} ~ = ~ \frac{F_{\sf X}}{M_{\rm Pl}}\: {\cal W}_{\Psi} \;+\; \frac{F_{\sf X}\: F_{\sf X}^\dagger}{ M_{\rm Pl}^2}\; K_{\Psi}
\end{equation}
and, in these conditions, they are completely universal.

\begin{figure}[t!]
    \begin{subfigure}[h]{\textwidth}
        \centering
        \includegraphics[width=0.9\textwidth]{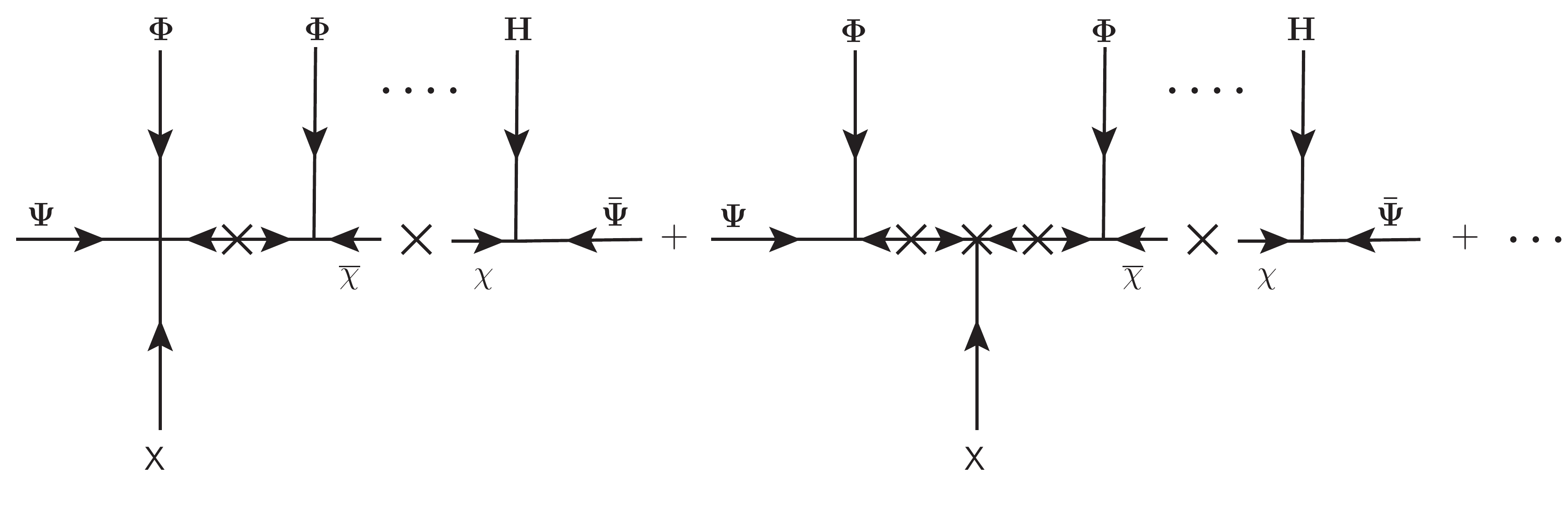}
        \subcaption{}
    \end{subfigure}
    \begin{subfigure}[h]{\textwidth}
        \hspace{0.4cm}
        \includegraphics[width=0.98\textwidth]{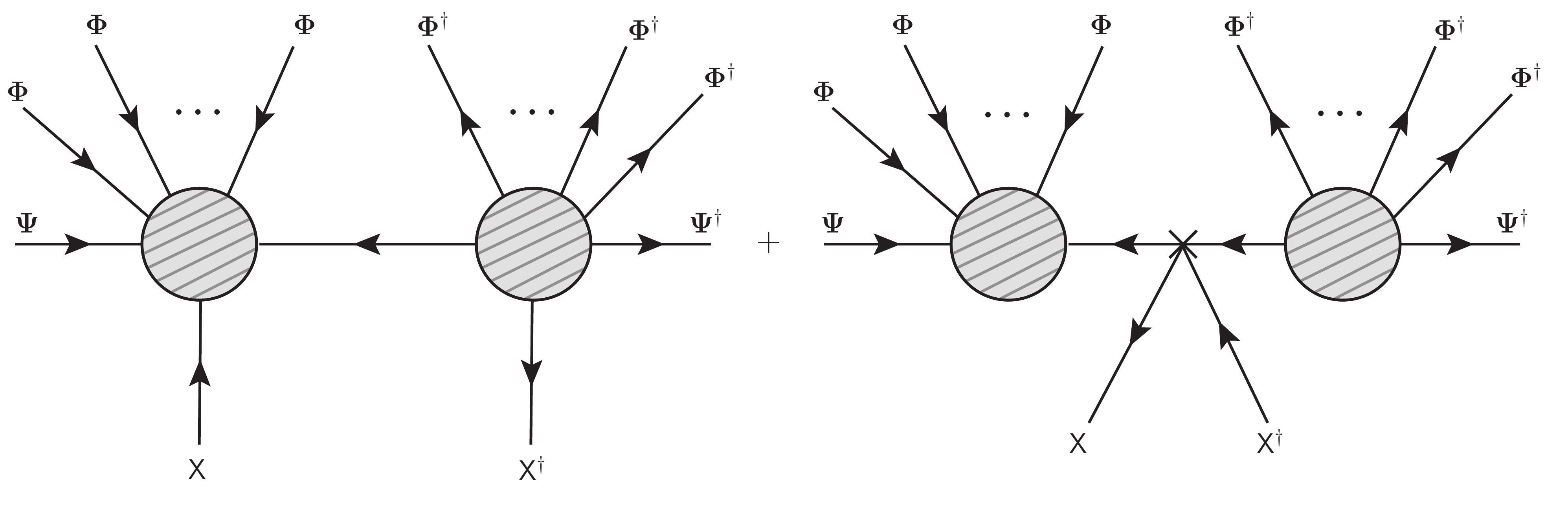}
        \subcaption{}
    \end{subfigure}
    \caption{\label{fig:superpot&trilin}
    \small 
    (a) A supergraph representation of two processes contributing to the same trilinear coupling.
    In the first diagram, the spurion field is attached to the vertex $\Psi$-$\Phi$-$\chi$ while, in the second diagram, it is inserted in the $\chi$ mass insertion.
    The same can be done for each flavon and Higgs insertion.
    Therefore, for a Yukawa diagram involving $n_{\rm in}$ flavons, there will be $(2\,n_{\rm in}+1)$ possibilities to generate the associated trilinear coupling.
    All of them have to  be taken into account.\\
    (b) A schematic supergraph depiction of the two types of processes that contribute to the same soft mass due to the insertion of the spurion ${\sf X X}^\dagger$ combination in different positions.
    The bubbles with $n_{\rm in}$ ($n_{\rm out}$) flavons and a $\sf X$ (${\sf X}^\dagger$) insertion symbolize the $(2\,n_{\rm in (out)}-1)$ possible diagrams resulting from attaching $\sf X$ as in the upper diagram.
    The second diagram accounts for the insertion of ${\sf X X}^\dagger$ in the internal heavy mediator.
    Combining all of them, $\big[(2\,n_{\rm in}-1)\,(2\,n_{\rm out}-1)+1\big]$ possibilities contribute to the soft mass.
    }
\end{figure}

However, after the breaking of the flavor symmetry, the superpotential and Kähler potential receive corrections from non-renormalizable operators coming from diagrams like those depicted in Figure \ref{fig:W&K_flavor}.
Integrating over the heavy mediators $\chi$, the Superpotential and Kähler potential can be schematically written as:
\begin{equation} \label{eqn:superpot}
    {\cal W}_{\Psi} \:=\: {\cal W}_{\Psi}^{\rm (ren)} \:+\: {\Psi\, \bar\Psi\, H}\; \sum_{\Phi}\sum_{n_{\rm in}=1}^{\infty}\, x_{{\rm n_{in}}}\, \left(\frac{ \langle \Phi \rangle}{M_\chi}\right)^{n_{\rm in}},
\end{equation}
\begin{equation} \label{eqn:kahler}
    K_{\Psi} \:=\: {\Psi}\, {\Psi}^\dagger\: \left[\, \mathbb{1} \:+\: \sum_{\Phi,\Phi^\dagger}\sum_{\substack{n_{\rm in},\\ \quad n_{\rm out}=1}}^{\infty}\, c_{({\rm n_{in},n_{out} })}\, \left(\frac{ \langle\Phi\rangle}{M_\chi}\right)^{n_{\rm in}}\left(\frac{ \langle\Phi^\dagger\rangle}{M_\chi}\right)^{n_{\rm out}}\,\right],
\end{equation}
where $\Psi$ stands for any field belonging to the visible sector\footnote{We follow the usual superfield notation throughout the article.
Thus, all the fields in Eqs.\eqref{eqn:superpot} and \eqref{eqn:kahler} must be read as superfields.
In Table \ref{tab:FieldTrans}, the relevant particle content for our analysis is specified.} and {\s ${\cal W}_\Psi^{\rm (ren)}$} consists of renormalizable operators such as the Higgs $\mu$-term or a possible top Yukawa coupling.
Therefore, the standard Yukawa couplings are only generated by these operators as powers of the expansion parameter $\epsilon=\langle \Phi \rangle/M_\chi$, where $\langle \Phi \rangle$ is the vev of the scalar flavon and $M_\chi$ the mass of the heavy mediators.
The bubbles of fields entering and leaving in Figure \ref{fig:W&K_flavor}-(b) are the sets of $n_{\rm in}$- and $n_{\rm out}$-flavon insertions in Eq.\eqref{eqn:kahler} which enclose similar structures to those in Figure \ref{fig:W&K_flavor}-(a).\\
Adding the spurion field to these non-renormalizable Yukawa couplings or to the K\"ahler potential generates the trilinear terms and soft mass matrices.
From Figure \ref{fig:superpot&trilin} one may see that, for each operator with $n$-flavon insertions, the number of effective operators contributing to the trilinear couplings and soft masses at the same order is equal to the number of different ways in which the spurion field can be inserted in the diagram.
Thus, the proportionality factor between trilinears (soft masses) and the corresponding Yukawa coupling (Kähler term) is given by:
\begin{align}
    A_{ij} & =~ (2\, n_{\rm in} + 1)\, a_0\, Y_{ij}, \label{eqn:Aij} \\[2pt]
    \left(\widetilde m_\Psi^2\right)_{ij} & =~ m_0^2\: f_{ij} \left(K_{\Psi}\right)_{ij} \hspace{0.5cm} {\rm with} \hspace{0.5cm} 
    f_{ij}=[\,(2\, n_{\rm in} -1)(2\, n_{\rm out}-1) + 1\,], \label{eqn:mij}
\end{align}
where $m_0\equiv \langle F_{\sf X} \rangle/M_{\rm Pl}$ and {$a_0\equiv k\, m_0$} with $k\sim \mathcal{O}(1)$.
The main consequence of Eqs. \eqref{eqn:Aij} and \eqref{eqn:mij} is that, although term by term the trilinears and the soft masses are proportional to Yukawas and Kähler elements, the full matrices are not proportional to them.
Therefore, going to the canonical basis, where the Kähler metric is the identity, and to the mass basis, where the Yukawas are diagonal, does not ensure that the soft terms are diagonalized.
Actually, as we will see below in the case of $A_5$, off-diagonal contributions will generally survive the rotations and they will have an impact on the low-energy phenomenology.

\section{\boldmath Lepton masses and mixing from $A_5$ and CP}
\label{sec:reviewA5CP}
$A_5$ is the non Abelian discrete group composed of the even permutations of five objects. It has $60$ elements and five irreducible representations (irrep): one singlet \1, two triplets \3 and $\tr$, one tetraplet \4 and one pentaplet \5. It can be generated by two elements, $s$ and $t$, satisfying\footnote{Lowercase letters are for the abstract elements of the group. Capital letters refer to specific representations in terms of $n\times n$ matrices.}:
\beq
    s^2 \,=\, (s\, t)^3 \,=\, t^5 \,=\, e.
\eeq
The specific form of these generators for each irrep of the group in our basis convention is shown in Appendix \ref{app:a5xCPgroup}.
The $A_5$ group contains several subgroups: fifteen associated with a $Z_2$ symmetry, five Klein subgroups $Z_2 \times Z_2$, ten related to $Z_3$ transformations and six $Z_5$ subgroups.
Combinations of them may play the role of residual symmetries for charged leptons and neutrinos.
Here we are interested in the combination of $A_5$ and CP as proposed in \cite{Feruglio:2012cw} (see also \cite{Holthausen:2012dk, Chen:2014tpa, Grimus:1995zi}), where the CP transformation generally acts non trivially on the flavor space \cite{Ecker:1983hz, Ecker:1987qp,Neufeld:1987wa}.
The action of a generalized CP transformation, $X$, over a field $\psi (x)$ is given by\footnote{Do not mistake the spurion chiral field responsible for the breaking of SUSY, $\sf X$, for the matrix representation of the generalized CP transformation X.}:
\beq
    \psi(x) ~\longrightarrow~ \psi'(x)\:=\: X \psi^\star (x_{\rm CP}),
\eeq
with $X$ a matrix representation of the CP transformation and $x_{\rm CP}=(x^0,-\overrightarrow x)$.
The transformation $X$ can be chosen as a constant, unitary and symmetric matrix:
\beq
    X X^\dagger \:=\: X X^\star \:=\: \mathbb{1}
\eeq
To ensure a consistent definition of the CP symmetry with the flavor group, the following condition must be verified:
\beq \label{eqn:A5timesCP} (X^{-1}\, A\, X)^\star \:=\: A' \eeq
where $A,\, A' \in A_5$.
In particular for $A_5$, Eq.\eqref{eqn:A5timesCP} is fulfilled with $A=A'$.\\
$A_5$ as a family symmetry for leptons leads to the Golden Ratio (GR) mixing, which predicts a vanishing reactor angle $\theta_{13}$.
A consequence of introducing CP as a symmetry is that the pure GR mixing is modified and a continuous parameter, \Oth, that quantifies this departure is introduced.
In fact, the small value of the reactor angle can be reproduced in terms of this variable that, at the same time, determines the amount of observable CP violation in the leptonic mixings.\\
The set of combinations of residual symmetries for {\s $A_5$ and CP} that accommodates well the observed mixing in the leptonic sector has been discussed in previous works \cite{DiIura:2015kfa, Li:2015jxa, Ballett:2015wia}.
Assuming that the lepton {\s $SU(2)_L$}-doublet, $\ell$, transforms like a triplet representation of $A_5$, the authors in \cite{DiIura:2015kfa} conclude that only four possibilities are allowed: two for {\s $G_e=Z_5$}, one for {\s $G_e=Z_3$} and another for {\s $G_e=Z_2\times Z_2$}; for neutrinos, {\s $G_\nu=Z_2\times$CP} has been always considered.
Here we are interested in the phenomenology of Case II of \cite{DiIura:2015kfa}, corresponding with {\s $G_e=Z_5$}.
The neutrino spectrum for this scenario has been fully analyzed in \cite{DiIura:2018fnk}.
\marisa{A} tuple of generators, $Q$ of ${\cal G}_e=Z_5$ and $(Z,\, X)$ of ${\cal G}_\nu=Z_2\times$CP, characterizing this realization is:
	\beq \label{eqn:tuple} (Q,\, Z,\, X)=(T,\, T^2ST^3ST^2, X_0), \eeq
where $X_0$ expressed in the \3 representation is
\beq \label{eqn:P23}
		X_0 \;=\; P_{23} \;\equiv\; \begin{pmatrix}
									~1~ & ~0~ & ~0~ \\
									 0  &  0  &  1  \\
									 0  &  1  &  0
        		  \end{pmatrix}.
\eeq
Its form in the rest of irreps of $A_5$ can be found in Appendix \ref{app:a5xCPgroup}.

\subsection{Charged-lepton masses}
\label{subsec:chargedmasses}

{\tablinesep=7pt
\begin{table}[t!]
\centering 
\begin{tabular}{|c|c c c c c c c c|}
\hline
~{\bf Mechanism} & $\ell$  & $\nu^c$ & $H_u$  & $\phi^\nu_{_{\bf 1}}$ & $\phi^\nu_{_{\bf 3}}$ & $\phi^\nu_{_{\bf     3'}}$ & $\phi^\nu_{_{\bf 4}}$ & $\phi^\nu_{_{\bf 5}}$ \\ \hline \hline
~ I       & ${\bf 3}$ & $-$        & ${\bf 1}$ & ${\bf 1}$  & $-$       & $-$ & $-$ & ${\bf 5}$ \\ \hline 
~ II $a$-2  & ${\bf 3}$ & ${\bf 3}$  & ${\bf 1}$ & ${\bf 1}$  & ${\bf 3}$ & $-$ & $-$ & ${\bf 5}$ \\ \hline
~ II $c$-2  & ${\bf 3}$ & ${\bf 3'}$ & ${\bf 1}$ & $-$ & $-$  & $-$       & ${\bf 4}$ & ${\bf 5}$ \\ \hline
\end{tabular}
\caption{\small \label{tab:FieldTrans}
Particle content for the different mechanisms examined here and its representation under the flavor group $A_5$ and CP.
Note that all these fields should be understood as chiral superfields that contains a spin-0 and a spin-1/2 component.}
\end{table}}

\begin{figure}[t!]
	\centering
	\includegraphics[scale=0.5]{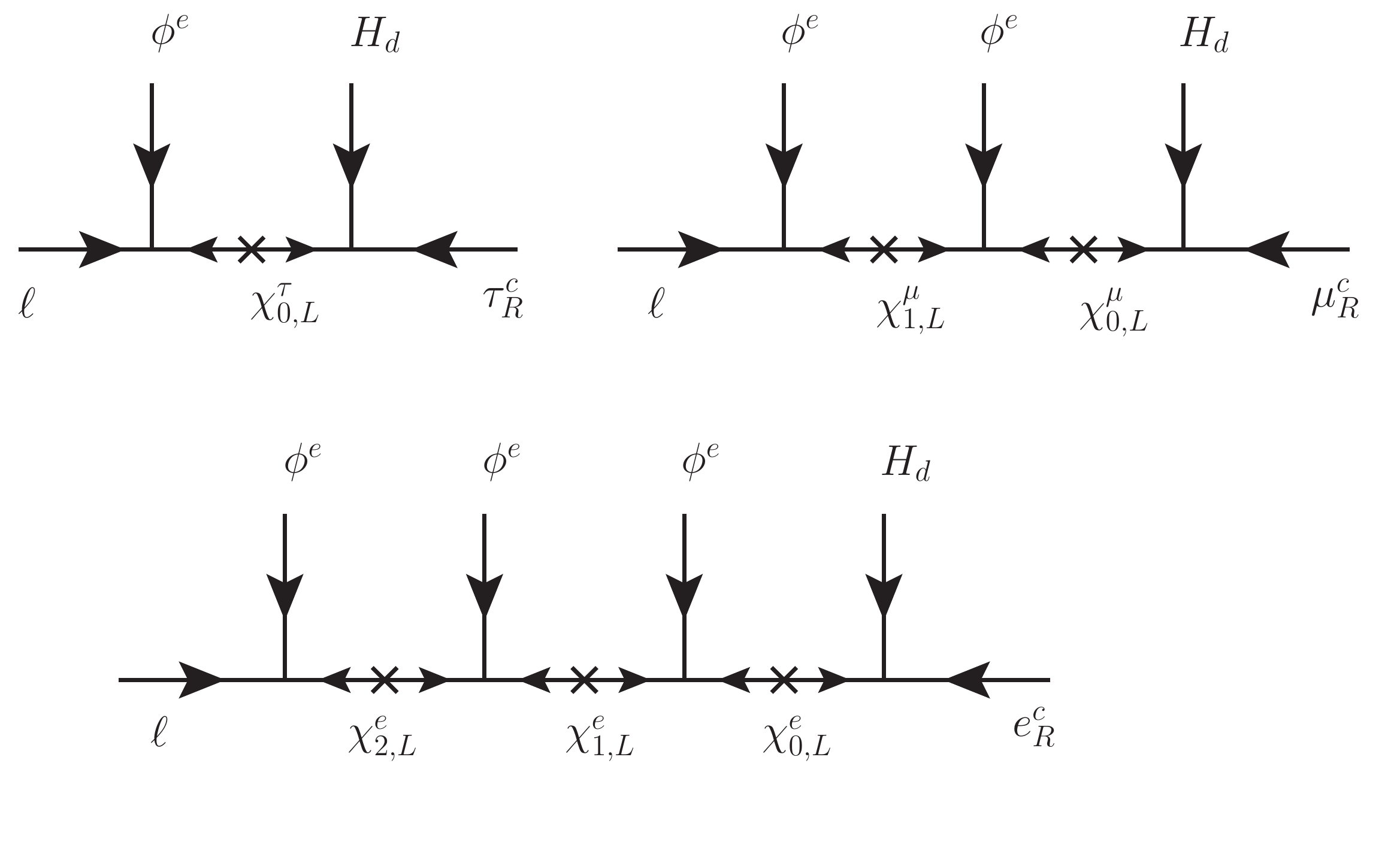}
	\caption{\label{fig:CLcase1} \small
Example where the hierarchy in the charged-lepton sector is due to the number of flavon insertions: tau, muon and electron masses are generated by diagrams involving one, two and three flavons, respectively. 
In this specific example the heavy mediators are doublets under $SU(2)_L$.}
\end{figure}

Residual symmetries constrain the form of the flavon vevs that break the invariance under the family symmetry ${\cal G}_\ell$.
Assuming that charged leptons are symmetric under the subgroup $G_e=Z_5$,
	\beq \label{eqn:feCond} Q_\ir\, \langle \phi^e_\ir \rangle=\langle \phi^e_\ir \rangle, \eeq
with the generators $Q_\ir$ and the flavon fields $\phi^e_\ir$ in the \ir\, representation, must be satisfied.
The condition in Eq.\eqref{eqn:feCond} implies that non-zero vevs are possible only for the triplet and pentaplet representations and their form is forced to be\footnote{Here we just provide the form of the flavon vevs symmetric under the residual symmetry. In general, the scalar potential responsible for them, which we do not specify here, will require the presence of additional superfields.}
\bea \label{eqn:chargedVEV}
        \langle \phi^e_{\bf 3}\rangle = \left(\begin{array}{c} \omega_3 \\ 0 \\ 0 
        \end{array}\right),\qquad
        \langle \phi^e_{\bf 3'} \rangle = \left(\begin{array}{c} \omega_{3'} \\ 0 \\ 0 
        \end{array}\right),\qquad
        \langle \phi^e_{\bf 5} \rangle = \left(\begin{array}{c} \omega_{5} \\ 0 \\ 0 \\ 
        0 \\ 0 \end{array}\right),
\eea
where $\omega_3$, $\omega_{3'}$ and $\omega_5$ are real parameters.
The flavons in Eq.\eqref{eqn:chargedVEV} generate non-renormalizable operators that enter the superpotential and give rise to the charged lepton masses as discussed in Eq.\eqref{eqn:superpot},
\beq \label{eqn:CharSuperPot}
	{\cal W}_\ell \supseteq {\cal W}_e \;=\; {\cal W}_e^{\rm ren} \:+\: \ell\, e^c_R\, H_d\, \sum_{n=1}^\infty 
    x_n\, \left(\frac{\langle \phi^e \rangle}{M_\chi}\right)^n,
\eeq
where we have not assumed any specific representation for the right-handed fields.
The residual symmetry also imposes an invariance requisite under the ($3\times 3$) effective mass matrix:
	\beq Q_\3^\dagger\; m_e^\dagger\,m_e\; Q_\3  =  m_e^\dagger\,m_e, \label{eqn:meCond} \eeq
where the generators are given in the triplet representation.
A straightforward consequence of Eq.\eqref{eqn:meCond} is that $m_e^\dagger m_e$ must be diagonal in the basis where the set of generators of the group $\lbrace Q_\ir \rbrace$ are diagonal.
In our case, the generator of $G_e=Z_5$ in the tuple of Eq.\eqref{eqn:tuple} is $Q=T$, which is diagonal according to Eq.\eqref{eqn:genrep}.
Therefore, the operators in Eq.\eqref{eqn:CharSuperPot} must produce a diagonal mass matrix at leading order (LO) (higher order corrections are discussed in Section \ref{subsec:NLOWrongFlavons}).
Its elements should exhibit the correct hierarchy between generations:
\beq \label{eqn:MassesCL}
    m_e \;\propto\; \begin{pmatrix}
    					~\lambda_c^4~	&		0			&	 0	\\
                                  0		&  ~\lambda_c^2~	&	 0	\\
                                  0		&		0			&	~1~
							\end{pmatrix},
\eeq
where $\lambda_c=0.2257$ stands for the Cabbibo angle.
In order to keep our discussion as model-independent as possible, we do not suppose any specific mechanism responsible for this pattern.
However, for completeness, we mention some possibilities that have been already proposed in the literature.\\
One way to generate the observed structure of masses for charged leptons is through processes like those depicted in Figure \ref{fig:CLcase1}, where the effective mass of each generation involves a different number of flavon insertions.
Each of them is proportional to the expansion parameter {\s $\varepsilon \equiv \langle \phi^e \rangle/M_\chi \ll 1$} that, in this specific case, might be {\s $\varepsilon\propto \lambda_C^2$}.
This type of diagrams can be easily arranged with an Abelian symmetry (continuous or discrete) assigning adequate quantum numbers to the lepton and flavon fields.
Another strategy could be having a symmetry that is broken at separated scales.
In this scenario, a natural hierarchy between the flavon vevs is expected.
Notice that the $\chi$ heavy mediators involved in all these constructions may be generally left-handed (LH) or right-handed (RH) under $SU(2)_L$.
In the absence of further hypothesis, we naturally expect all of them to be present and have masses of the same order.

\begin{figure}[t!]
    \centering
	\includegraphics[scale=0.5]{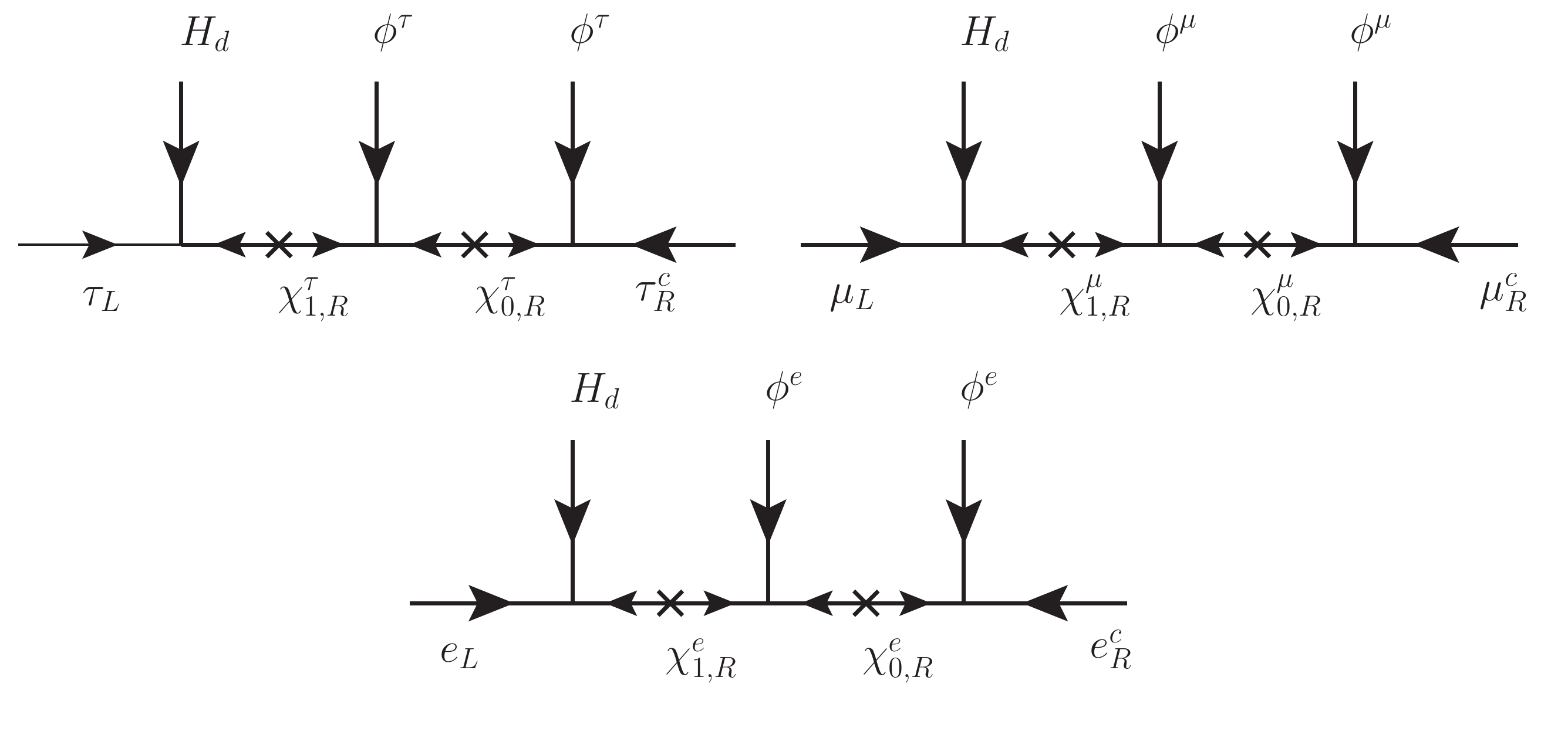}
	\caption{\label{fig:CLcase2} \s
Example where the hierarchy for the charged-lepton masses comes from the flavon vevs that break the family symmetry at different scales.
Then, {\footnotesize $\langle \phi^e \rangle \ll \langle \phi^\mu \rangle \ll \langle \phi^\tau \rangle$} is expected.
In this specific example the heavy mediators are singlets under $SU(2)_L$.}
\end{figure}

\subsection{Neutrino masses}
\label{subsec:neutrinos}
The residual symmetry in the neutrino sector, ${\cal G}_\nu=Z_2\times$CP, delimits the form of the light neutrino mass matrix $m_\nu$ through the following invariance conditions:
\bea \label{eqn:MassInvNeu}
	Z^T_\3\; m_\nu\; Z_\3 \:=\: m_\nu \hspace{0.5cm} & {\rm and} & \hspace{0.5cm} X_\3\; m_\nu\; X_\3 \:=\: m_\nu^*, 
\eea
where $Z_\3$ and $X_\3$ are the generators of $Z_2$ and CP in the triplet representation.
The texture of the matrix that satisfies Eq.\eqref{eqn:MassInvNeu} is
\beq \label{eqn:NeutMass}
	m_\nu = m^\nu_0 \begin{pmatrix}
				s + x + z & \dfrac{3}{2\sqrt{2}}(z + i \varphi y) & \dfrac{3}{2\sqrt{2}}(z - i \varphi y)\\
				\dfrac{3}{2\sqrt{2}}(z + i \varphi y) & \dfrac{3}{2} (x + i y) & s - \dfrac{x + z}{2} \\
				\dfrac{3}{2\sqrt{2}}(z - i \varphi y)  & s - \dfrac{x + z}{2} & \dfrac{3}{2} (x - i y) 
				\end{pmatrix},
\eeq
where $\lbrace s,\, x,\, y,\, z \rbrace$ are dimensionless and real numbers, {\s $\varphi=(1+\sqrt{5})/2$} is the GR and $m^\nu_0$ is the absolute mass scale for neutrinos.
Similarly, the flavon vevs breaking the family symmetry in this sector must satisfy
\bea \label{eqn:FlavInvNeu}
	Z_\ir\, \langle \phi^\nu_{\bf r} \rangle \:=\: \langle \phi^\nu_{\bf r} \rangle \hspace{0.5cm} & {\rm and} & \hspace{0.5cm}
	X_\ir\, \langle \phi^\nu_{\bf r}\rangle^* \:=\: \langle \phi^\nu_{\bf r} \rangle,
\eea
where $\langle \phi^\nu_\ir \rangle$ and the generators of $Z_2$ and CP, $Z_\ir$ and $X_\ir$, are  expressed in the \ir\, representation.
The vacuum alignment is then subject to verify the general structure\footnote{As for the charged leptons, we simply present the structure of the flavon vevs compatible with the residual symmetry in the neutrino sector.
The scalar potential originating them, which we do not compute here, will generally involve additional superfields to those detailed in Table \ref{tab:FieldTrans}.}:
\bea
	\langle \phi^\nu_{\bf 1} \rangle \,=\, \upsilon_1,\qquad 
    \langle \phi^\nu_{\bf 3}\rangle\,=\, \upsilon_3\, 
    										 \left(\begin{array}{c} -\sqrt{2}\varphi^{-1} \\
    																					1 \\
                                                                                        1 	
    												\end{array}\right), \qquad
    \langle \phi^\nu_{\bf 3'} \rangle = \upsilon_{3'}\,
    									  \left(\begin{array}{c} \sqrt{2} \varphi \\
                                          										1 \\
                                                                                1 
    											\end{array}\right)\,, \label{eqn:neutralVEV1}
\eea
\bea
	\langle \phi^\nu_{\bf 4} \rangle \,=\, \left(\begin{array}{c} y_r -i y_i \\
    												   (1+2\varphi)y_r - i y_i \\
                                                       (1+2\varphi) yr + i y_i \\
                                                                    yr + i y_i
													\end{array}\right), \qquad
	\langle \phi^\nu_{\bf 5} \rangle = \left(\begin{array}{c} -\sqrt{\frac{2}{3}}(x_r + x_{r,2}) \\
    																			-x_r+i \varphi x_i \\ 
                                                                               		 x_{r,2}-i x_i \\
                                                                                     x_{r,2}+i x_i \\
                                                                                 x_r+i \varphi x_i 		
												\end{array}\right), \label{eqn:neutralVEV2}
\eea
where all the coefficients are real.\\
Majorana neutrino masses can be generated through the so-called dimension $5$ Weinberg operator, which is produced at tree level by type I, II or III see-saw mechanisms.
Each case is related to the addition of one extra particle to the SM spectrum: RH neutrinos, a scalar triplet or a fermion triplet, respectively.
In the effective low-energy theory some of these constructions are equivalent to others (a detailed discussion about this can be found in Appendix C of \cite{DiIura:2018fnk}) so that we can reduce the discussion to just two cases: Mechanism I, consisting of the Weinberg operator and type II see-saw, and Mechanism II, which includes type I and III see-saw realizations.
In the following, we specify the operators generating the neutrino masses for each mechanism.
The quantum numbers of the fields for each scenario are displayed in Table \ref{tab:FieldTrans}, where the nomenclature of \cite{DiIura:2018fnk} has been conserved.

\subsubsection*{Mechanism I}
\label{subsubsec:mecI}
\begin{figure}[t!]
  \centering
  \includegraphics[scale=0.5]{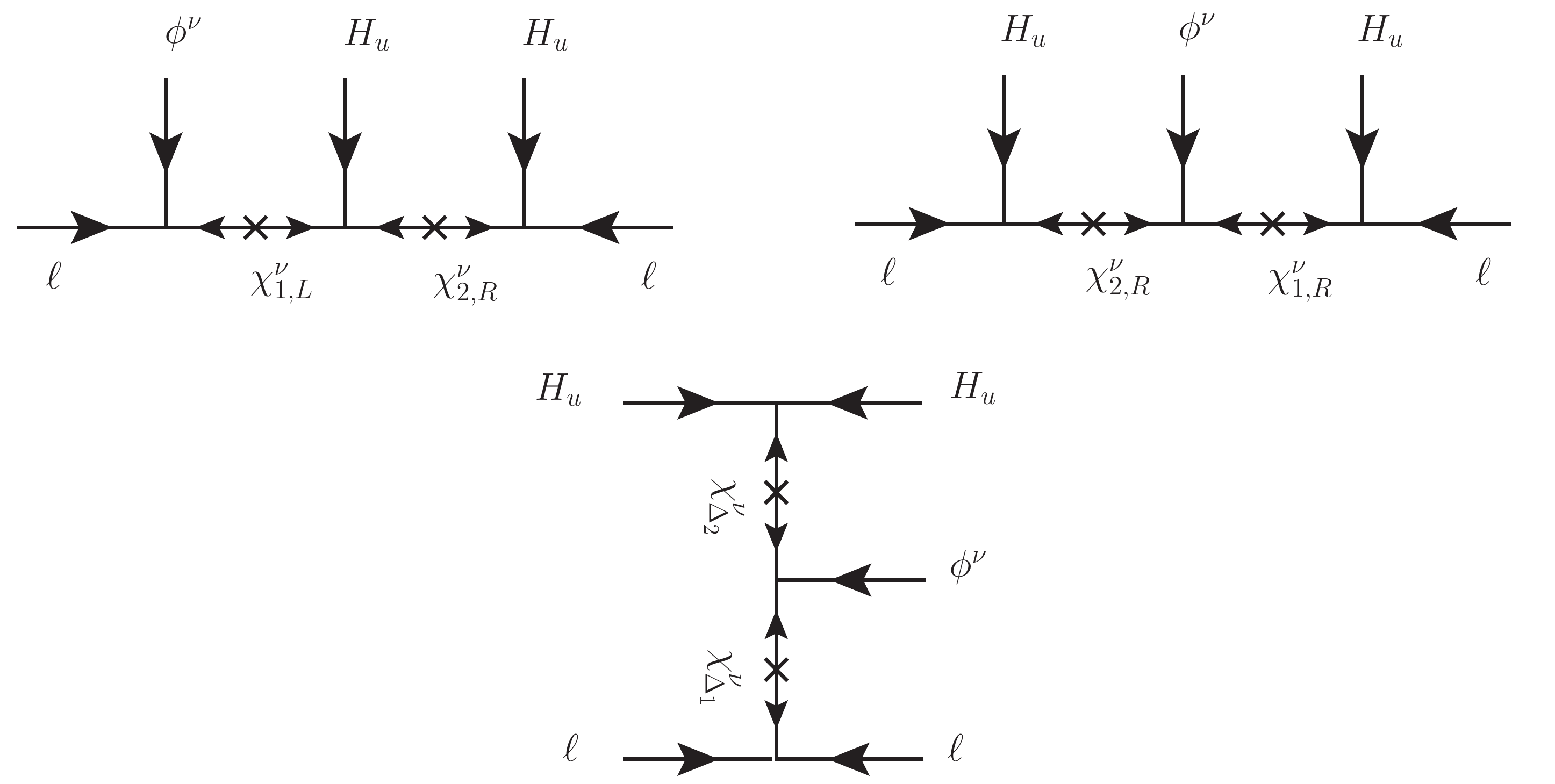}
  \caption{\label{fig:mecIneut} \small
Different contractions generating the Weinberg Operator of Mechanism I in Section \ref{subsubsec:mecI}.
The first diagram involves $SU(2)_L$ doublets and singlets as mediators whereas the second and third only require singlets or triplets.
The phenomenological implications in each case are different. 
}
\end{figure}

Here we consider neutrino masses generated by the Weinberg operator.
In this case, a lepton doublet $\ell$ transforming as the \3 or the $\tr$\, representation produces the same phenomenological results under a redefinition of the parameters as indicated in \cite{DiIura:2018fnk}.
Therefore, without loss of generality, one may simply consider $\ell\sim \3$.
The LO contributions to the effective superpotential responsible for the neutrino masses are:
\beq
	{\cal W}_\ell \supset {\cal W}_{\nu}^{\rm I} \;=\; y_1^\nu\, \frac{\left[\ell^2 H_u^2\, \phi_{\bf 
    1}^\nu\right]_{\bf 1}}{M_\chi^2} \;+\; y_5^\nu\, \frac{\left[\ell^2 H_u^2\, \phi_{\bf 5}^\nu\right]_{\bf 1}}
    {M_\chi^2},
\eeq
where $y_i^\nu$ are dimensionless parameters and brackets mean that different contractions are possible.
In the low-energy theory all of them are equivalent, since they give rise to the same predictions.
However, as can be seen in Figure \ref{fig:mecIneut}, the mediator sector involved in each process is very different and thereby their ultraviolet (UV) origin: while in the first diagram the heavy messengers are both LH and RH, the second process considers RH fields and the last one scalar triplets.
In the absence of additional hypothesis about the high-energy theory, all these mediators are present and have similar masses, $M_\chi$.\\
Once the flavor symmetry is broken, the mass matrix in Eq.\eqref{eqn:NeutMass} is generated with
\beq \label{eqn:MecIparam}
	s\,\propto\, y^\nu_1\frac{\upsilon_1}{M_\chi} \hspace{0.5cm}
    x\,\propto\, -y^\nu_5\frac{x_{r,2}}{M_\chi} \hspace{0.5cm}
    y\,\propto\, -y^\nu_5\frac{x_i}{M_\chi} \hspace{0.5cm}
    z\,\propto\, -y^\nu_5\frac{x_r}{M_\chi},
\eeq
where $\upsilon_1$ and $\lbrace x_{i},\, x_{r},\, x_{r,2}\rbrace$ are the vevs of the singlet and pentaplet flavons respectively, see Eqs.\eqref{eqn:neutralVEV1} and \eqref{eqn:neutralVEV2}.

\begin{figure}[h!]
  \centering
  \vspace{1.cm}
  \includegraphics[scale=0.5]{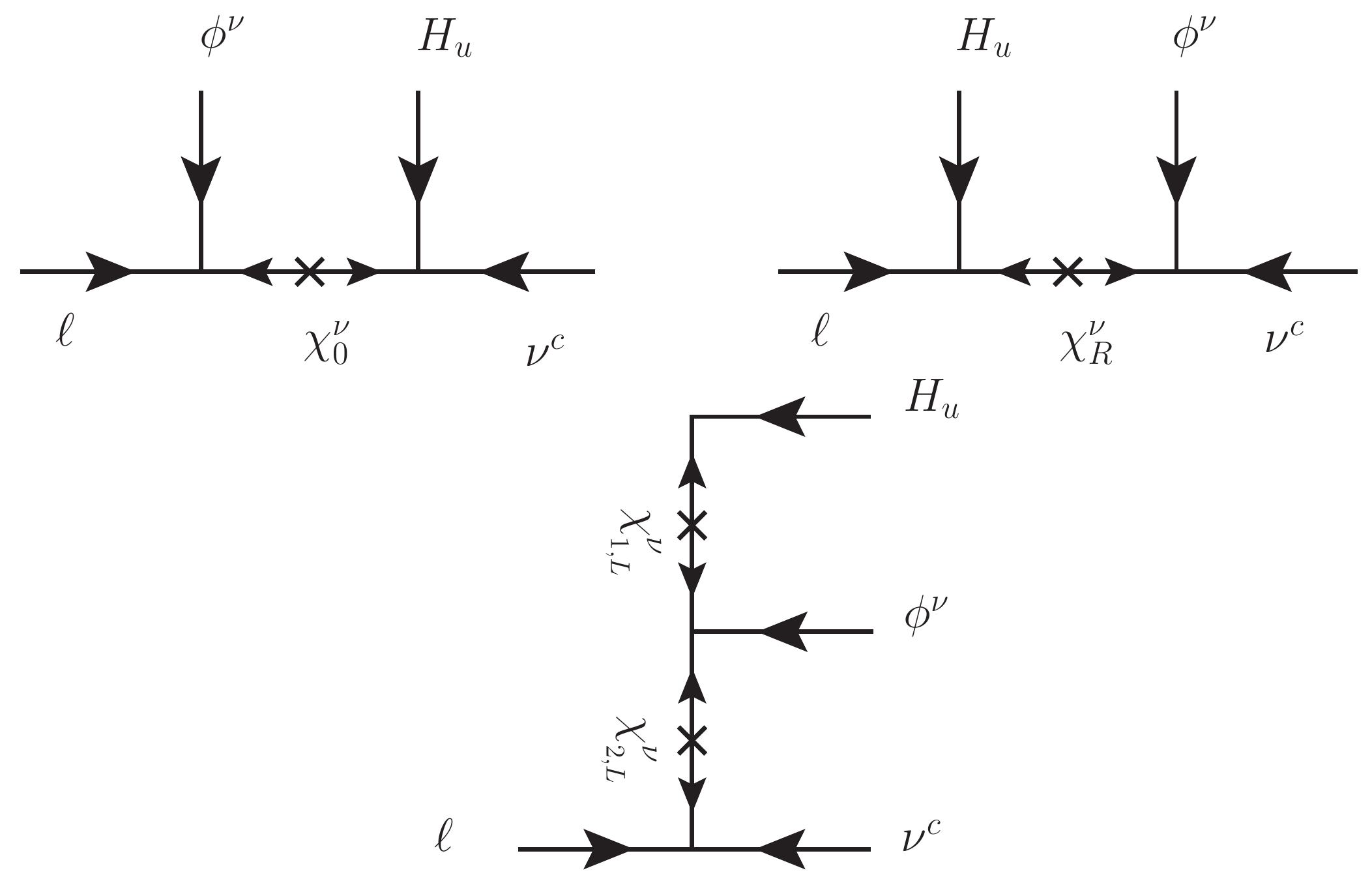}
  \caption{\label{fig:mecIIneut} \small 
Different contractions generating the Dirac mass matrix of Mechanism II $a$-2 and $c$-2 in Section \ref{subsubsec:mecII}.
The first and third diagrams involve $SU(2)_L$ doublets as mediators whereas the second requires singlets.
The phenomenological implications in each case are different.
}
\end{figure}

\subsubsection*{Mechanism II}
\label{subsubsec:mecII}
The type I see-saw formulation is investigated in Mechanism II.
Two different situations are analyzed in this framework: both $\ell$ and $\nu^c$ transform in the same triplet representation (II $a$-2) or in opposite ones (II $c$-2).
The specific choice, $\ell\sim \3$ or $\ell \sim \tr$, does not affect the resulting phenomenology.
Therefore, $\ell \sim \3$ is always assumed.
Here we consider a trivial Majorana matrix, namely {\s $M_M=M P_{23}$}, with $P_{23}$ defined in Eq.\eqref{eqn:P23}, while the Dirac mass matrix, {\s $M_D$}, is more complicated.
The opposite option, non-trivial Majorana mass and trivial Dirac matrix, does not introduce observable effects in the LFV observables examined in the following section, hence we do not discuss this possibility further.\\
The effective operators entering the superpotential that produces the Dirac mass are:
{\s \beq \label{eqn:WMecIIa2}
	{\cal W}_{\nu}^{{\rm II}a2} \;=\; Y_1^\nu\, \left[\, \nu^c \ell \,\right]_{\bf 1} H_u \;+\; 
    y_1^\nu\,\frac{\left[\, \nu^c \ell\, \phi_{\bf 1}^\nu \,\right]_{\bf 1} H_u}{\Lambda} \;+\; y_3^\nu\, \frac{\left[\, \nu^c \ell\, 
    \phi_{\bf 3}^\nu \,\right]_{\bf 1} H_u}{\Lambda} \;+\; y_5^\nu\, \frac{\left[\, \nu^c \ell\, \phi_{\bf 5}^\nu \,\right]_{\bf 1} H_u}
    {\Lambda} \;+\; {\rm c.c.}
\eeq}
for Mechanism II $a$-2, and
{\s \beq \label{eqn:WMecIIc2}
	{\cal W}_\ell \supset {\cal W}^{{\rm II}c2}_\nu \;=\; y_4^\nu\, \frac{\left[\, \nu^c \ell\, \phi_{\bf 4}^\nu \,\right]_{\bf 1} H_u}
    {\Lambda} \;+\; y_5^\nu\, \frac{\left[\, \nu^c \ell\, \phi_{\bf 5}^\nu \,\right]_{\bf 1} H_u}{\Lambda} \;+\; {\rm c.c.}
\eeq}

for Mechanism II $c$-2.
Again the brackets indicate that different contractions are possible.
Figure \ref{fig:mecIIneut} shows the UV origin of each of them involving LH mediators, RH mediators and triplets under $SU(2)_L$. 
Once the flavor symmetry is broken, the neutrino mass matrix is generated through the usual type I see-saw process,
	\beq \label{eqn:seesawI} m_\nu \;=\; -M_D^T\cdot M_M^{-1}\cdot M_D \;=\; -\frac{1}{M}\, M_D^T\cdot P_{23}^{-1}\cdot M_D. \eeq
The coefficients $\lbrace s,\, x,\, y,\, z \rbrace$ entering $m_\nu$ as in Eq.\eqref{eqn:NeutMass} are intricate linear combinations of the dimensionless parameters:
\beq \label{eqn:MecIIa2param} 
	      f\,\propto\,Y_1 \,+\, y^\nu_1\frac{\upsilon_1}{M_\chi} \hspace{0.5cm}
          g\,\propto\,y^\nu_3\frac{\upsilon}{M_\chi} \hspace{0.5cm}
        h_r\,\propto\,y^\nu_5\frac{x_r}{M_\chi} \hspace{0.5cm}
		h_i\,\propto\,y^\nu_5\frac{x_i}{M_\chi} \hspace{0.5cm}        
    h_{r,2}\,\propto\,y^\nu_5\frac{x_{r,2}}{M_\chi}
\eeq
for Mechanism II $a$-2, and
\beq \label{eqn:MecIIc2param} 
	    f_r\,\propto\,y^\nu_4\frac{y_r}{M_\chi} \hspace{0.5cm}
		f_i\,\propto\,y^\nu_4\frac{y_i}{M_\chi} \hspace{0.5cm}
        h_r\,\propto\,y^\nu_5\frac{x_r}{M_\chi} \hspace{0.5cm}
		h_i\,\propto\,y^\nu_5\frac{x_i}{M_\chi} \hspace{0.5cm}        
    h_{r,2}\,\propto\,y^\nu_5\frac{x_{r,2}}{M_\chi}
\eeq
for Mechanism II $c$-2.

\subsection{Lepton Mixing}
\label{subsec:MixingA5xCP}
The \upmns is the matrix that measures the misalignment between the rotations of LH charged-leptons and neutrinos to the mass basis.
It is defined as
	\beq \label{eqn:upmnsdef} U_{\rm PMNS} \;=\; U_{e}^\dagger\, U_\nu\,. \eeq
The conservation of CP and residual transformations determine the form of this matrix up to permutations of rows and columns.
In our case, the residual symmetry in the charged sector indicates that for ${\cal G}_e=Z_5$ generated by $Q=T$ diagonal, the unitary rotation to the mass basis for LH charged leptons {$U_e$} is the identity.
Therefore the \upmns will be determined only by the neutrino mixing at LO.
The unitary transformation that diagonalizes the neutrino mass matrix in Eq.\eqref{eqn:NeutMass} is \cite{Feruglio:2012cw}
	\beq \label{eqn:upmns} U_{\rm PMNS} =  \Omega_\nu\, R^{13}_\theta\, K_\nu\,, \eeq
where $\Omega_\nu$ defines a change of basis that block-diagonalizes the initial matrix in Eq.\eqref{eqn:NeutMass} and it is directly related to the GR mixing, $R_\theta^{13}$ is a rotation of an angle \Oth in the $1-3$ plane and $K_\nu$ is a diagonal matrix with entries $\lbrace \pm 1,\, \pm i \rbrace$ needed to have positive eigenvalues.
The explicit form of the matrix $\Omega_\nu$ is
\beq \label{eqn:OmegaMatrix}
    \Omega_\nu = \frac{1}{\sqrt{2}}\begin{pmatrix}
							\sqrt{2}\,\cos\phi & \sqrt{2}\,\sin\phi & \quad 0 ~\\
							\sin\phi & -\cos\phi & \quad i ~\\
							\sin\phi & -\cos\phi & \;-i ~ \end{pmatrix},
\eeq
with {\s $\sin\phi \equiv 1/\sqrt{2 + \varphi}$} and {\s $\cos\phi \equiv \sqrt{(1 + \varphi)/(2 + \varphi)}$}.
The rotation is given by
\beq \label{eqn:R13Matrix}
	R^{13}_\theta \;=\; \left(\begin{matrix}
					~ \cos\theta ~ & ~ 0 ~ & ~ \sin\theta ~ \\
						      0    &   1   &   0			\\
					   -\sin\theta &   0   & \cos\theta \end{matrix} \right),
\eeq
where the size of the angle \Oth is totally fixed by the entries of the block-diagonalized mass matrix.
In our case:
\beq \label{eqn:tan2O}
	\tan 2\theta \;=\; \frac{2\sqrt{7 + 11 \varphi}\, y}{2 x(\varphi +1) \:+\: z(2\varphi+1)}.
\eeq
\\
Mixing angles and complex phases can be directly extracted from Eqs.\eqref{eqn:upmns}$-$\eqref{eqn:tan2O} using the standard {\s \upmns}parametrization that we detail in Appendix \ref{app:upmns}.
The reactor angle is proportional to the angle $\theta$ as:
	\beq \label{eqn:sin2O13} \sin^2\theta_{13} \;=\; \frac{2+\varphi}{5}\, \sin^2\theta. \eeq
Eq.\eqref{eqn:sin2O13} implies that a small value of $\theta$ is required in order to reproduce $\theta_{13}\sim 9^\circ$.
Indeed, $\theta_{\rm bf}=0.175$ has been found as the best fit value for this realization in \cite{DiIura:2015kfa}.
Inspecting Eq.\eqref{eqn:tan2O}, one may see that such a tiny value for $\theta$ can only be obtained considering the following hierarchy among vevs\footnote{A natural way of obtaining this hierarchy is through the two step symmetry breaking ${\cal G}_\ell \rightarrow {\cal G}_\nu=Z_2\times Z_2\times$CP $\rightarrow Z_2\times$CP as explained in \cite{DiIura:2018fnk}.}: $y\ll x, z, s$.\\
The atmospheric angle is predicted to be maximal, {\s $\sin^2\theta_{23}=1/2$}, and the solar angle is related to the reactor angle through the sum rule:
\beq \label{eqn:sinO12}
	\sin^2\theta_{12} \;=\; \frac{3-\varphi}{5\cos^2\theta_{13}} \;\simeq\; \frac{0.276}{\cos^2\theta_{13}}
\eeq
CP invariants and complex phases are also predicted in this framework.
The Jarlskog invariant \cite{PhysRevLett.55.1039} is
	\beq \label{eqn:JInv} J_{\rm CP} \;=\; \frac{1}{8}\sin2\theta_{12}\sin\theta_{23}\sin2\theta_{13}\cos\theta_{13}\sin\delta \;=\;-\frac{\sqrt{2+\varphi}}{20}\, \sin2\theta. \eeq
The Dirac phase $\delta$ can be inferred from Eq.\eqref{eqn:JInv} and it is maximal, {\s $|\sin\delta|=1$}.
The other CP invariants, defined as
\bea
	I_1 & \equiv & {\rm Im}{\bigg[U_{12} U_{12}U^*_{11}U^*_{11}\bigg]} = \sin^2\theta_{12}
	\cos^2\theta_{12}\cos^4\theta_{13}\sin\alpha \label{eqn:CPInvI1} \\
	I_2 & \equiv & {\rm Im}{\bigg[U_{13} U_{13}U^*_{11}U^*_{11}\bigg]} = \sin^2\theta_{13}
	\cos^2\theta_{12}\cos^2\theta_{13}\sin\beta,	\label{eqn:CPInvI2}
\eea
vanish exactly.
Hence Majorana phases, $\alpha$ and $\beta$, must be $0$ or $\pi$.

\subsection{Next-to-Leading Order Corrections}
\label{subsec:NLOWrongFlavons}
Notice that Eqs.\eqref{eqn:chargedVEV} and \eqref{eqn:neutralVEV1}-\eqref{eqn:neutralVEV2} indicate that the LO masses of the neutral and charged leptons must be induced by two separate set of flavons.
In practice, this division can be always ensured introducing an additional $Z_N$ symmetry that distinguishes among them at LO\footnote{
For instance, considering the diagrams in Figure \ref{fig:CLcase2} and the first diagram in Figure \ref{fig:mecIIneut} as the responsible mechanisms for lepton masses, one may see that the charges discriminating both sectors would be:
\begin{eqnarray}
	N_{\chi^l_n} & = & N_{H_l}+N_{l^c}+n N_{\phi_l}-(n+1)N\,, \nn \\
	N_{\phi^e} & = & (3N-N_\ell-N_{H_d}-N_e^c)/2 \,, \nn \\
    N_{\phi^\nu} & = & 2N-N_\ell-N_{H_u}-N_\nu^c\,, \nn
\end{eqnarray}
with $l\equiv\{e,\mu,\tau,\nu\}$.
For the remaining cases, similar assignments can be done.}.
However, at higher orders, flavons belonging to the opposite sector (the {\it wrong} flavons) are allowed to enter the LO diagrams and may introduce sizable corrections.
In this section we comment how these effects can be adequately taken into account.\\
The NLO corrections to the leading predictions of masses and mixing 
usually come from higher-order operators that enter the scalar potential and the superpotential. 
The former induce a shift in the flavon vevs while the latter is usually generated by extra flavon insertions to the LO operators.
As commented before, some of these insertions may be due to the {\it wrong}-flavons.
For neutrinos, one may check that these contributions are usually subleading.
Assuming that the heavy fields $\chi$ mediating the diagrams in both sectors have masses of the same order, corrections to the neutrino mass matrix are of the form
	\beq \label{eqn:NLOneutrinos} \delta m_\nu \sim \lambda_C^2\: m'_\nu, \eeq
where $m'_\nu$ is a {\s $3\times 3$} matrix with elements given in terms of the $\lbrace s,\, x,\, y,\, z \rbrace$ parameters defined in Eq.\eqref{eqn:NeutMass}.
Therefore, barring accidental cancellations in the LO elements which may make the NLO terms dominant, we expect these corrections to be mostly subleading.\\
Similarly, a suitable choice of charge assignments under a $Z_N$ symmetry can guarantee that insertions of the neutrino flavons to the charged sector are also subleading.
For instance, if the following set of charges are considered
\beq \label{eqn:flavonsZN} 
	N_{\phi^\tau} = N_{\phi^\mu} = N_{\phi^e} = k \quad {\rm and} \quad N_{\phi^\nu}=1,
\eeq
such that ${\cal Q}_{Z_N}(\phi)=e^{i\,2\pi N_{\phi}/N}$, it is easy to see that one charged-lepton flavon could only be substituted by $k$ neutrino-flavon insertions\footnote{
For the diagrams in Figures \ref{fig:CLcase2} and \ref{fig:mecIneut} (Mechanism I), the charges of the flavons in Eq.\eqref{eqn:flavonsZN} can be easily obtained solving the equation system:
\begin{eqnarray}
	N_{\phi^\nu} = & N-2\,N_\ell-2\,N_{H_u} & = 1\,, \nn \\
	N_{\phi^\tau} = & N_{\rm CL}-N_{\tau^c} & = k\,, \nn \\
	N_{\phi^\mu} = & (N_{\rm CL}-N_{\mu^c})/2 & = k\,, \nn \\
	N_{\phi^e} = & (N_{\rm CL}-N_{e^c})/3 & = k, \nn
\end{eqnarray}
with $N_{\rm CL}=N-N_{H_d}-N_\ell$. For other cases, one may proceed in a similar way.}.
Then, in its most general form, corrections to the charged-lepton mass matrix will be given by
\beq 
	\delta m_e \sim \left(\frac{\langle\phi^\nu\rangle}{M_\chi}\right)^k \begin{pmatrix}
    				~1~ & ~1~ & ~1~\\
                     1  &  1  & ~1~\\
                     1  &  1  & ~1~\\ \end{pmatrix}.
\eeq
Adjusting the value of $k$, the required suppression can always be obtained.

\section{Kähler Potential and Soft Terms}
\label{sec:KahlerSoft}
In Section \ref{sec:reviewSUSY}, the main consequences of embedding a flavor symmetry in supersymmetric theories where $\Lambda_{\rm SUSY} \gg \Lambda_{f}$ have been discussed, see also \cite{Das:2016czs, Lopez-Ibanez:2017xxw, deMedeirosVarzielas:2018vab, Calibbi:2012yj, Antusch:2011sq}.
Even in the most conservative case where SUSY breaking is parametrized by a single universal spurion field, {\it tree-level} flavor violating effects generally arise from the mismatch between the order one coefficients in the soft-breaking structures and the Yukawa and kinetic terms due to the different equivalent options of inserting this spurion field \cite{Das:2016czs}.
Moreover, we observe that, in the absence of further hypothesis over the UV spectrum, the common origin of the flavor structures for charged leptons and neutrinos may induce testable relations between observables belonging to these two sectors.
Finally, we notice that, for some configurations, significant corrections from the Kähler potential emerge that break the residual symmetries and could modify the LO predictions.
This fact is independent of SUSY and it actually corresponds to the usual wave-function renormalization that should always be taken into account \cite{King:2003xq, King:2004tx, Espinosa:2004ya, Chen:2013aya,Antusch:2011sq}.

\subsection{Kähler corrections}
\label{subsec:KahlerCorr}
The effective Kähler potential for a multiplet $\Psi$ of $A_5$ and CP can be schematically written as
\beq \label{eqn:Kahler}
	K_\Psi \;=\; \Psi^\dagger_i \Psi_j \left[~\delta_{ij} + \sum_{n,\,a,\,b}\; c_{ij\, ab}^n  \left( 
    \frac{\langle\phi^{l\,\dagger}_{\ir_a}\rangle\; \langle\phi^l_{\ir_b}\rangle}{M_\chi}\right)^n~\right],
\eeq
with $l=e,\nu$ and $M_\chi$ the generic mass of the $\chi$ heavy mediators.
The specific terms entering the sum depend on the symmetries of the model.
We are interested in the Kähler potential for lepton doublets and singlets.
Under no further assumptions, we can at least write the following terms for the former: 
\bea \label{eqn:KahlerLeft}
	K_\ell & = & \left[ 1 + \frac{1}{M_\chi^2}\, \sum_{\ir} \left( \phi^{e\dagger}_\ir\; \phi^e_\ir \,+\, 
    \phi^{\nu\dagger}_\ir\;\phi^\nu_\ir \right) \;+\; \dots \right] \times \nn \\
	& & \left[ \ell^\dagger \ell  \;+\; \frac{1}{\Lambda^2} \sum_{\ir} \left(\big[\,\ell^\dagger \phi^{e\dagger}_\ir\,\big]_\1\, 
    \big[\,\ell \phi^e_\ir\,\big]_\1 \,+\, \big[\,\ell^\dagger \phi^{\nu\dagger}_\ir\,\big]_\1\, 
    \big[\,\ell \phi^\nu_\ir\,\big]_\1 \right) \;+\; \dots \right], \label{eqn:KahlerLHex}
\eea
where dots stand for higher order terms in the same fashion and the sum in \ir\, accounts for the flavons at work depending on the mechanism, see Table \ref{tab:FieldTrans}. 
The first line is associated with singlet contractions that contribute to every element of the Kähler metric, so they can be factorized as a global constant.
The second line corresponds to non-trivial contractions and generates off-diagonal entries in the Kähler metric.
Note that, being $\phi \phi^\dagger$ combinations, all terms in Eq.\eqref{eqn:KahlerLeft} are neutral under any possible charge. Hence, they cannot be avoided and must be adequately taken into account.
The Kähler function for RH charged-leptons can be written as:
{\s \bea
	K_R^e & = & \left( 1 + \frac{1}{M_\chi^2}\, \sum_{\forall \ir} \phi^{e\dagger}_\ir\;\phi^e_\ir \;+\; \cdots \right)
    			\times \left( e_R^{c\, \dagger} e_R^c  \;+\; \frac{1}{M_\chi^2} \sum_{\forall \ir} \big[\,e_R^{c\, \dagger} 
                \phi^{e\dagger}_\ir\,\big]_\1\, \big[\,e_R^c \phi^e_\ir\,\big]_\1 \;+\; \cdots \right).
                \label{eqn:KahlerRightE}                
\eea}

In contrast to the previous case, here only the flavons associated with charged leptons contribute.
An analogous expression can be written for RH neutrinos replacing {\s $\phi^e \to \phi^\nu$} and {\s $e_R^c \to \nu^c$}.\\
The Kähler structures in Eqs.\eqref{eqn:KahlerLeft} and \eqref{eqn:KahlerRightE} can be explicitly computed simply by inserting the flavon vevs in Eqs.\eqref{eqn:chargedVEV} and \eqref{eqn:neutralVEV1}-\eqref{eqn:neutralVEV2}.
The results are rather intricate, so we omit them here.
We find that the charged-lepton flavons only contribute to the diagonal of the Kähler metric, while neutrino flavons also induce off-diagonal entries.
Because of that, a redefinition of the fields is required in order to go to the physical basis where the kinetic terms are canonical, that is $K^c_\Psi=\mathbb{1}$.
This can be always achieved making use of an upper triangular matrix that decomposes the Hermitian Kähler metric as $K_\Psi=U_c^\dagger U_c$ \cite{King:2004tx}:
\beq \label{eqn:CN} 
    \Psi^c \quad\longrightarrow\quad \Psi^c\;=\; U_c\, \Psi ~/\quad \Psi^\dagger\,K_\Psi\,\Psi = \Psi^{c\,\dagger}\,\Psi^c
\eeq
The upper triangular matrix defined by Eq.\eqref{eqn:CN} has the schematic form
\beq \label{eqn:UppTriang}
	U_c \;\simeq\; \begin{pmatrix}
    			~1+\varepsilon_{11}^2~	&    \varepsilon^2_{12}  &   \varepsilon^2_{13}   \\
                    		0			& ~1+\varepsilon^2_{22}~ &   \varepsilon^2_{23}   \\
                    		0			&   		 0 			 & ~1+\varepsilon^2_{33}~ \\
                    \end{pmatrix},
\eeq
where $\varepsilon_{ij}^2$ are complex entries linearly dependent on the flavon vevs squared.
In our case, the size of these entries may vary from {\s $10^{-7}$} up to {\s $10^{-2}$}, depending on the mechanism\footnote{Notice that this upper-triangular form ensures that the corrections of canonical normalization are always subleading for hierarchical matrices.}.
The field redefinition to the canonical basis affects the lepton masses in Eqs.\eqref{eqn:MassesCL} and \eqref{eqn:NeutMass}, which should be rotated as:
\bea
	m_e & \longrightarrow & m_e^c\;=\;\left(U_c^{-1}\right)^\dagger \,m_e\, U_c^{-1} \,, \label{eqn:CNCLmass} \\
    m_\nu & \longrightarrow & m_\nu^c\;=\;\left(U_c^{-1}\right)^T \,m_\nu\, U_c^{-1} \,. \label{eqn:CNNeutrmass}
\eea
Initially, the charged leptons in the non-canonically normalized flavor basis satisfy that $m_e^\dagger\,m_e$ is diagonal and hierarchical.
The canonical rotation introduces corrections to the product in Eq.\eqref{eqn:MassesCL} as:
\beq \label{eqn:CNMassesCL} 
	m_e^\dagger\, m_e \quad\longrightarrow\quad m_e^{c\, \dagger}m_e^c \;\simeq\; \begin{pmatrix}
							~\lambda_c^8~	  & ~\varepsilon_{12}^2\,\lambda_c^8~ & ~\varepsilon_{13}^2\,\lambda_c^8~ \\
          		    					\cdot &     	     \lambda_c^4	  	  & ~\varepsilon_{23}^2\,\lambda_c^4  \\
                   						\cdot & 						  \cdot   & 				1 	  \end{pmatrix}.
\eeq
The corrected diagonalizing matrix at LO is then
\beq \label{eqn:CNVeL} U_e \quad\longrightarrow\quad U_e^c \;=\; U_e \,+\, \delta U_e \;\simeq\; 
							\begin{pmatrix}
							1 & ~\varepsilon_{12}^{2^\star}\, \lambda_c^4~ & 
                            ~\varepsilon_{13}^{2^\star}\,\lambda_c^8~\\
                            ~-\varepsilon_{12}^2\, \lambda_c^4~ & ~1 & \lambda_c^8 \\
                            ~-\varepsilon_{13}^2\,\lambda_c^8 & -\lambda_c^8 & 1
    						\end{pmatrix}.
\eeq
It enters the new \upmns but does not introduce sizable modifications to the initial angles.
This is in agreement with \cite{King:2004tx}, where the authors analyzed the effects of the canonical rotation for hierarchical Yukawa matrices.\\
The situation for neutrinos is slightly more involved because their mass matrix is not completely hierarchical and analytical expressions accounting for these effects are difficult to obtain.
We expect that corrections from the canonical rotation are important when affecting small quantities, such as the reactor angle or the lightest mass for a very hierarchical spectrum, and in those cases where the neutrino spectrum has two or more quasi-degenerated masses, since small contributions can modify the ordering.
Numerically, this has been confirmed: while the correlations among vevs obtained in \cite{DiIura:2018fnk} approximately remain after the canonical rotation, the range of possible values for the free parameters is significantly reduced because of these effects\footnote{Notice, however, that these corrections will affect only the determination of the parameters of a given model from the experimental results.
Almost always these variations lead to no measurable effects on the observable predictions.}.

\subsection{Soft-breaking Masses}
\label{subsec:SoftMasses}
Once we have the Kähler potential, the soft-breaking masses before canonical normalization can be obtained just examining the effective operators in $K_\Psi$, as detailed in Section \ref{sec:reviewSUSY}.
At LO, the soft-mass matrices are proportional, element by element, to the Kähler metric, see Eq.\eqref{eqn:mij}, and the proportionality factor, $f_{ij}$, accounts for the different ways in which the spurion F-term $F_{\sf X}$ can be inserted in the representative Kähler diagram \cite{Das:2016czs, Lopez-Ibanez:2017xxw}.
In our case:
\bea 
	\widetilde m_{\ell}^2 & = & m_{3/2}^2 \left[ \ell^\dagger \ell  \;+\; \frac{2}{M_\chi^2} \sum_{\ir} \left(\big[\,\ell^\dagger 
    	\phi^{e\dagger}_\ir\,\big]_\1\, \big[\,\ell \phi^e_\ir\,\big]_\1 \,+\, \big[\,\ell^\dagger \phi^{\nu\dagger}_\ir\,\big]_\1\, 
    	\big[\,\ell \phi^\nu_\ir\,\big]_\1 \right) \;+\; \cdots \right],\label{eqn:SoftMassLH} \\
	\widetilde m_{e_R}^2 & = & m_{3/2}^2 \left[ e_R^{c\, \dagger} e_R^c  \;+\; \frac{2}{M_\chi^2} \sum_{\forall \ir} \big[\,e_R^{c\, 
    	\dagger}\phi^{e\dagger}_\ir\,\big]_\1\, \big[\,e_R^c \phi^e_\ir\,\big]_\1 \;+\; \cdots \right]. 
        \label{eqn:SoftMassRH}   
\eea
Going to the canonical basis,
\bea \label{eqn:msoftCN}
	\widetilde m_\Psi^2 \quad\longrightarrow\quad \widetilde m^{c\, 2}_\Psi \;=\; 
    \left(U_c^{-1}\right)^\dagger\; \widetilde m^2_\Psi\; U_c^{-1},
\eea
and then to the mass basis, 
\bea \label{eqn:msoftUPMNS}
	\widetilde m_\Psi^{c\,2} \quad\longrightarrow\quad U_{\rm PMNS}^{c \dagger}\; \widetilde m^{c\,2}_\Psi\; U_{\rm PMNS}^c\,, 
\eea
one observes that some off-diagonal terms in the LH soft mass matrix of Eq.\eqref{eqn:msoftUPMNS} survive the rotations.
They can manifest in LFV processes such as $l_j \to l_i\,\gamma$, $l_j \to 3\,l_i$ or $\mu - e$ conversion.
Moreover, since the LFV contributions for LH charged-sleptons and the neutrino flavor structure arise from the same flavons, testable relations between observables from each sector can be inferred.
On the other hand, RH charged leptons do not receive observable corrections to their soft masses since they are generated by the flavons in Eq.\eqref{eqn:chargedVEV}, which only contribute to the diagonal.
As previously discussed for the Kähler potential, they can simply be reabsorbed through a redefinition of the fields.

\section{Predictions on flavor observables}
\label{sec:results}
Here we perform a combined analysis of the phenomenology of charged leptons and neutrinos for the two mechanisms exposed in Section \ref{subsec:neutrinos}.
Following the strategy of \cite{DiIura:2018fnk}, each case is divided in several subcases, where one (Mechanism I) of two (Mechanism II) vevs of the active neutrino flavons are set to zero\footnote{In many situations, that is equivalent to leave out some of the flavons in a model; if not, it could be arranged working out the correspondent vacuum alignment that produces it.}.
This allows to reduce the number of independent parameters and to inspect relations among the vevs that correctly reproduce the neutrino properties showed in Table \ref{tab:neutrinoBounds}.
We make predictions for the total sum of the light neutrino masses, the effective mass $m_{\beta\beta}$ and the flavor changing processes collected in Table \ref{tab:LFVbounds}.
Our numerical scan is realized as follows:
\begin{itemize}
\item First, we randomly generate the independent parameters corresponding to each mechanism in a range $[-1,+1]$, Eqs.\eqref{eqn:MecIparam}, \eqref{eqn:MecIIa2param} and \eqref{eqn:MecIIc2param}.
Then, we compute the light neutrino mass matrix $m_{\nu}$ in Eq.\eqref{eqn:NeutMass}, the Kähler metric in Eq.\eqref{eqn:KahlerLeft} and the soft masses in Eq.\eqref{eqn:SoftMassLH}.
The mass matrix for charged leptons is assumed to be like in Eq.\eqref{eqn:MassesCL}.
\item The structures obtained before have to be rotated to the canonical basis as indicated in Eqs.\eqref{eqn:CNCLmass}, \eqref{eqn:CNNeutrmass} and \eqref{eqn:msoftCN}.
After that, the lepton mass matrices are diagonalized and re-phased to obtain real and positive eigenvalues.
\bea
	U_\nu^{c\,\dagger}\,m_\nu^{c\,\dagger}m_\nu^c\,U^c_\nu = {\rm 
    Diag}(m_1^2,m_2^2,m_3^2) \,,\\
	U_e^{c\, \dagger}\,m_e^{c\,\dagger}m_e^c\,U^c_e = {\rm Diag}
    (m_e^2,m_\mu^2,m_\tau^2)\,.
\eea
At this stage, the \upmns defined as in Eq.\eqref{eqn:upmnsdef}, with $U_l \to U^c_l$, and the mass splittings\footnote{We assume that RGEs effects for the neutrino parameters are negligible so that our observables computed at high scales, $\Lambda\simeq 10^{14}-10^{16}$ GeV, can be directly compared with the data in Table \ref{tab:neutrinoBounds}.
Although this is the case for a hierarchical neutrino spectrum, it is not necessarily true for degenerate masses \cite{Antusch:2003kp}.
However, in our case, the degenerate scenarios correspond to those subcases that are already disfavored by cosmological bounds.}, {\s $\Delta m_{21}^2=m_2^2-m_1^2$} and {\s $\Delta m_{3j}^2=m_3^2-m_j^2$} with $j=1$ for Normal Hierarchy (NH) and $j=2$ for Inverted Hierarchy (IH), are checked to be in the $3\sigma$-allowed region for both hierarchies, see Table \ref{tab:neutrinoBounds}.
\item For those points which correctly reproduce lepton masses and mixing in the $3\sigma$ region, we fix the value of $m_0^\nu$ comparing with the best fit result for $\Delta m_{21}^2$.
The value of $m_0^\nu$ is expected to be some eV at most, so we discard points that correspond to $m_0^\nu>5$ eV.
\item The soft mass matrices are evolved from the GUT to the EW scale by means of the MSSM renormalization group equations and the branching ratios for the lepton flavor violating observables in Table \ref{tab:Cosmobounds} are computed\footnote{Numerical calculations for the running, spectrum and low-energy processes have been performed through the Supersymmetric Phenomenology package (SPheno) \cite{Porod:2003um}, together with the SARAH Mathematica package \cite{Staub:2013tta} for the generation of the source code.}.
We also provide estimations for the total sum of light neutrino masses and the effective mass $m_{\beta\beta}$.
\end{itemize}

\subsection{Neutrinos}
\label{subsec:resultsneutrinos}
\begin{figure}
  \centering
  \includegraphics[scale=0.62]{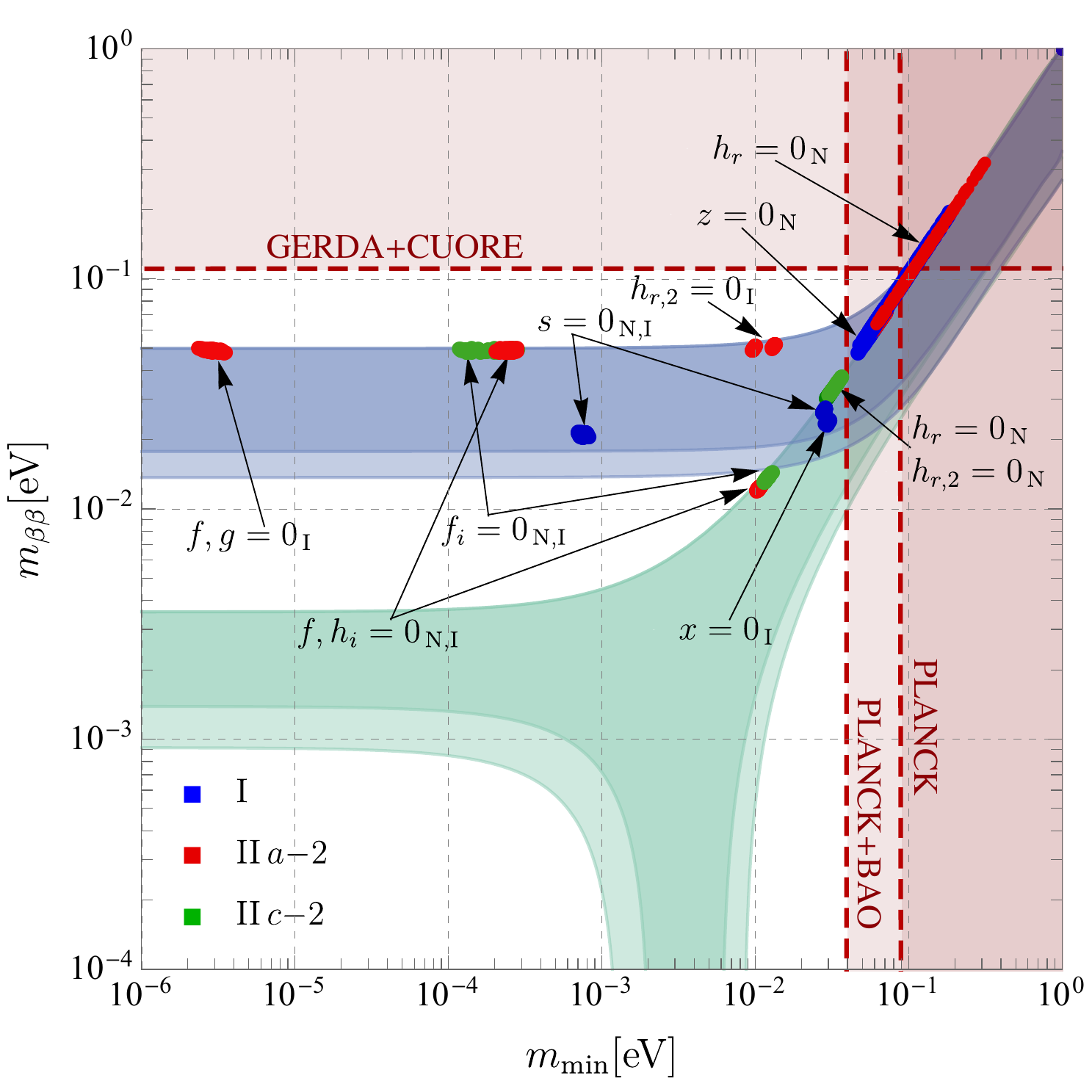}\\
  \includegraphics[scale=0.62]{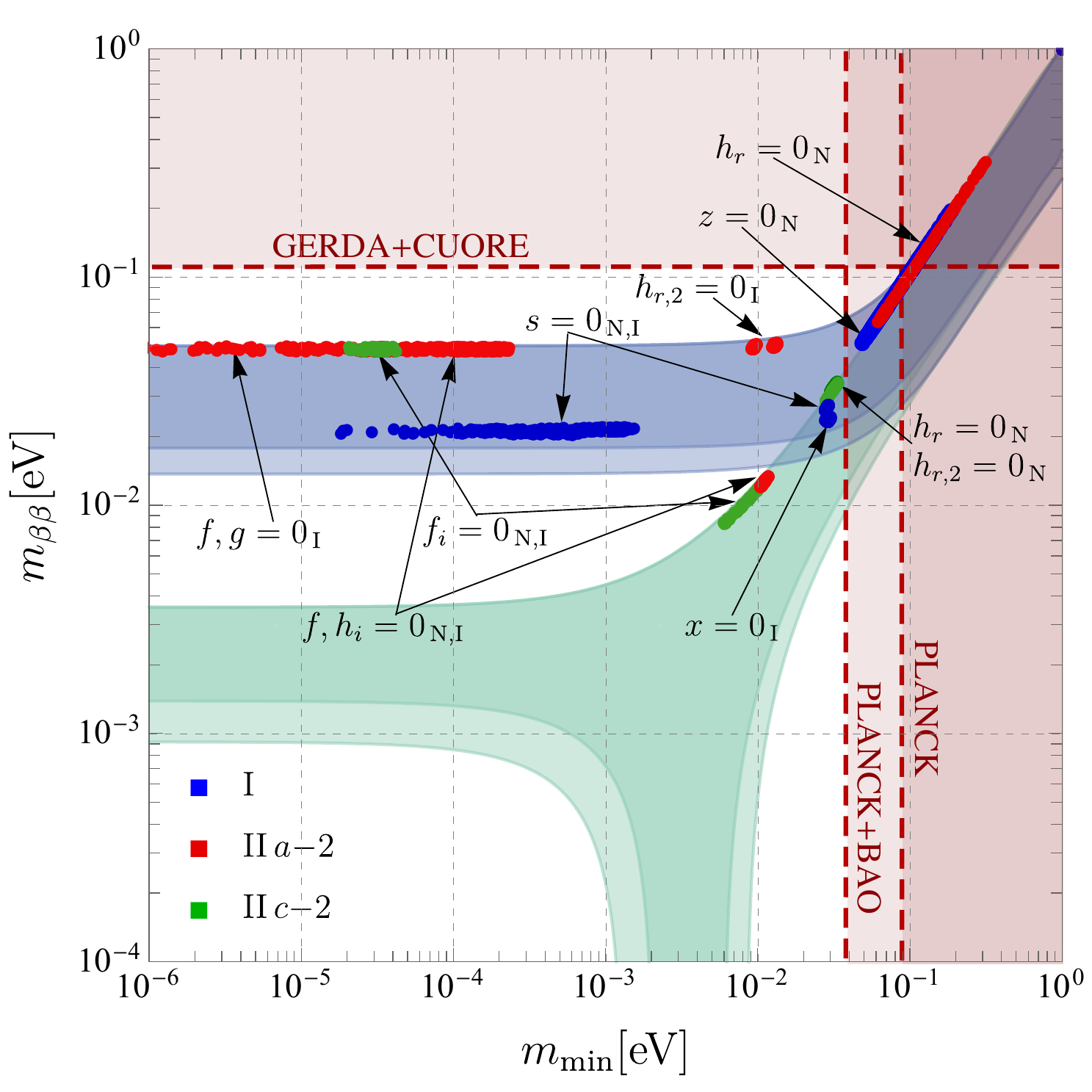}
  \caption{\s \label{fig:mbetabetaVSmin}
  The observable $m_{\beta\beta}$ against $m_{\rm min}$.
  The (bottom) top panel corresponds to the assumption of UV LH-mediators originating (non-)negligible corrections to the Kähler metric.
    Blue points correspond to Mech.\,I, red points to Mech.\,II $a$-2 and green points are due to Mech.\,II $c$-2.
    The green and blue shadows are the ones allowed for NO and IO respectively.
    Red shaded regions are already excluded by experimental observations.}
\end{figure}

For the neutrino sector, we follow the analysis performed in \cite{DiIura:2018fnk}.
The results are displayed in Figure \ref{fig:mbetabetaVSmin}, where the effective neutrino mass $m_{\beta\beta}$ is plotted against the minimum neutrino eigenvalue, that is $m_{\rm min}\equiv m_1$ for NH and $m_{\rm min}\equiv m_3$ for IH.
The upper panel reproduces the estimations presented in \cite{DiIura:2018fnk} whilst the lower panel shows the effect of rotating to the canonical basis where the K\"ahler is the identity.
Blue points are related to Mechanism I, where neutrino masses are produced by the Weinberg operator or type II see-saw.
Red points to Mechanism II $a$-2, corresponding to type I and III see-saw with trivial Majorana matrix and left- and right-handed neutrinos transforming in the same triplet representation.
Green points are associated with Mechanism II $c$-2, which considers the same framework as Mechanism II $a$-2 but with the lepton doublet and the neutrino singlet transforming in different triplet representations.
According to \cite{DiIura:2018fnk}, each case is divided in several subcases where one (Mechanism I) or two (Mechanism II) flavon vevs are set to zero.
Each subcase allows for NH, IH or both of them.
For Mechanism I, we have three subcases corresponding to $z=0$, $x=0$ and $s=0$.
For Mechanism II $a$-2, a total of six subcases are studied: $f=0$, $h_r=0$ and $h_{r,2}=0$ with either $h_i=0$ or $g=0$.
Finally, Mechanism II $c$-2 consists of six more cases: $f_i=0$, $h_r=0$ and $h_{r,2}=0$ with either $h_i=0$ or $f_r=0$.
For conciseness we identify the different cases with the first vev and specify the second vev only when necessary.
In addition, we use the subscripts "N" and "I" as shorthand notation for NH and IH.  \\
From Figure \ref{fig:mbetabetaVSmin} we conclude that the scenarios $z=0_{\rm\,N}$ of Mechanism I and $h_r=0_{\rm\,N}$ of Mechanism II $a$-2 are incompatible with the latest data from \texttt{Planck} + BAO, while other cases like Mechanism I with $s=0_{\rm\,N}$ and $x=0_{\rm\,I}$ or Mechanism II with $h_r=0_{\rm\,N}$ and $h_{r,2}=0_{\rm\,N}$ could be tested in the future with further cosmological data.
As shown in Table \ref{tab:Cosmobounds}, future sensitivity in neutrinoless double beta decay experiments could probe those realizations that predict IH and the region for quasi-degenerate masses with a sensitivity $m_{\beta\beta}=(0.01\div 0.05)$ eV (for reviews see \cite{Dell'Oro:2016dbc, deSalas:2018bym}).\\
Comparing upper and lower panels of Figure \ref{fig:mbetabetaVSmin}, it is evident that the predictions on $m_{\rm min}$ for some subcases change significantly and result in being extended by one or two orders of magnitude.
That happens for those hierarchical subcases where the smallness of $m_{\rm min}$ make it more susceptible to higher order corrections from the canonical rotation.
Significant cancellations between leading and higher order terms happen in Mechanism I with $s=0_{\rm\,I}$, Mechanism II $a$-2 with $f=0_{\rm\,I}$, and Mechanism II $c$-2 with $f_i=0_{\rm\,N}$, see the second plot of Figure \ref{fig:mbetabetaVSmin}.
However, note that a direct measurement of $m_{\rm\,min}$ is of no feasible pursue in the near future.
From this point of view, $m_{\rm\,min}$ can be considered as much as an additional model parameter so we may conclude that the canonical rotation has no substantial consequences on the testable predictions of the model in \cite{DiIura:2018fnk}.
The two exceptions to this are the subcases Mechanism II $a$-2 with $\lbrace f,\, h_i \rbrace=0_{\rm N}$ and Mechanism II $c$-2 $f_i=0_{\rm N}$ where, with enough resolution, a small discrepancy with the predictions in \cite{DiIura:2018fnk} for $m_{\beta\beta}$ may be experimentally found.

\subsection{Charged Leptons}
\label{subsec:resultsCL}
\begin{figure}
  \centering
  \includegraphics[scale=0.50]{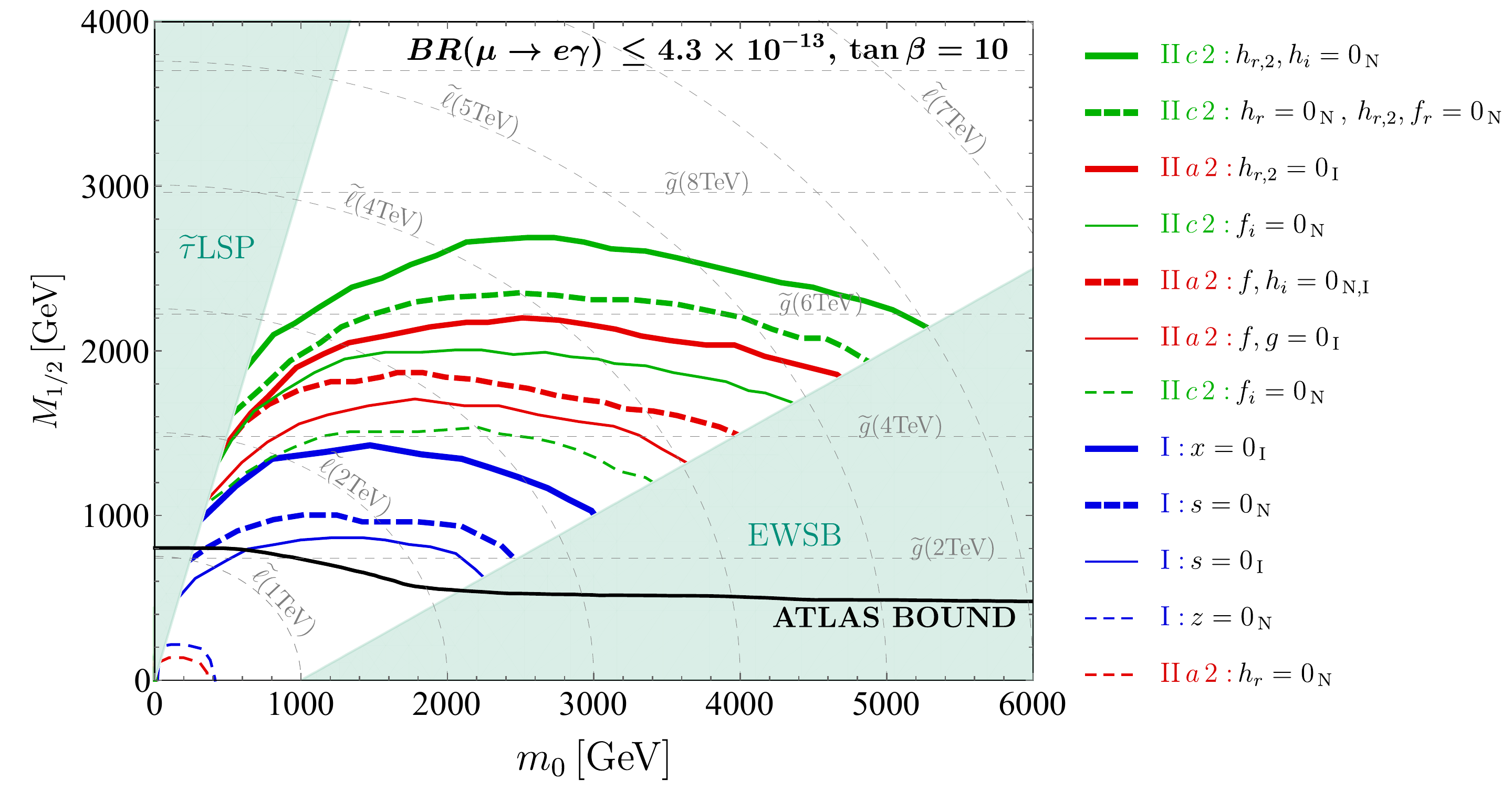}\\
  \includegraphics[scale=0.5]{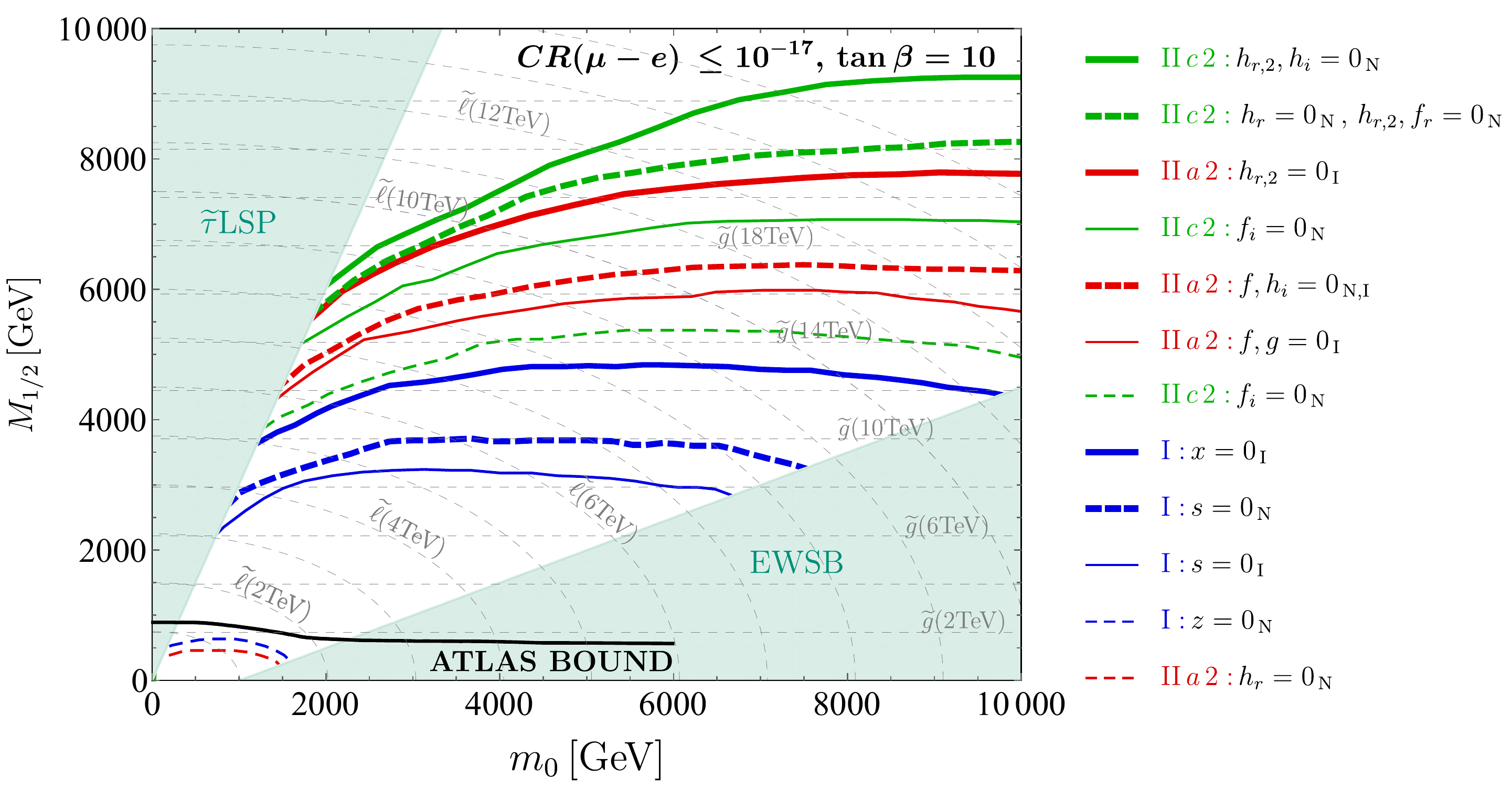}
  \caption{\label{fig:m0M12_BRmueyExcluded} \small 
Excluded regions of the MSSM parameter space due to LFV constraints for $\tan\beta=10$.
The upper panel is obtained imposing the current bound for ${\rm BR}(\mu\rightarrow e \gamma)$, the bottom panel show the regions expected to be ruled out if the future sensitivity for ${\rm CR}(\mu-e)_{\rm Al}$ is reached with no discovery.
The different mechanisms are distinguished through different colors: I (blue), II $a$-2 (red) and II $c$-2 (green).
The black line is the ATLAS bound for mSUGRA models.}
\end{figure}
It has been pointed out in previous works the importance of correctly account for the flavor effects that emerge from the inclusion of a flavor symmetry in SUSY \cite{Das:2016czs, Lopez-Ibanez:2017xxw, deMedeirosVarzielas:2018vab}.
On the one hand, it allows for a characterization of the flavor models of interest through their contributions to flavor-changing (FC) observables while, on the other hand, the search for experimental signals on those processes will help us to explore the SUSY spectrum at energies that go far beyond the LHC high-luminosity upgrade.
In this section we follow the strategy in \cite{Lopez-Ibanez:2017xxw} and we study the flavor contributions arising from the breaking of $A_5$ and CP in the LFV observables collected in Table \ref{tab:LFVbounds}.\\
The structure of the soft mass matrix for LH fields is given in Eq.\eqref{eqn:SoftMassLH}.
We compute its numerical value for each of the points that reproduce the neutrino experimental parameters in the $3\sigma$ region, Table \ref{tab:neutrinoBounds}, plus the \texttt{Planck} bound on the total sum of light neutrino masses, Table \ref{tab:Cosmobounds}.
The range of possible values for the off-diagonal entries results in being constrained by the phenomenology (upper bound) and by the absolute mass scale $m_0^\nu\sim 5$ eV (lower bound).
The interval corresponding to each case is reported in Table~\ref{tab:deltasRanges}.\\
The minimum value of the off-diagonal elements in Table~\ref{tab:deltasRanges} can be used to constrain the MSSM parameter space as it has been done in \cite{Das:2016czs, Lopez-Ibanez:2017xxw, deMedeirosVarzielas:2018vab}.
Concretely, we compute the branching ratio (BR) for the LFV processes displayed in Table \ref{tab:LFVbounds} considering an hypothetical soft mass matrix for LH sleptons whose off-diagonal elements correspond to those minimum values.
We choose a representative value of $\tan\beta=10$ and calculate our predictions for different values of $\lbrace m_0,\, M_{1/2} \rbrace$.
Comparing with current and future experimental bounds in Table \ref{tab:LFVbounds}, we are able to set excluded regions in the $\{m_0,M_{1/2}\}$ plane.
Barring accidental cancellations, we expect that bounds for specific models based on the realizations studied throughout this work will be like those presented in Figure \ref{fig:m0M12_BRmueyExcluded} in the most optimistic scenario.\\
The most interesting results are shown in Figure \ref{fig:m0M12_BRmueyExcluded}, where the upper panel has been obtained from the most restrictive process nowadays, BR$(\mu\rightarrow e \gamma) \leq 4.3 \times 10^{-13}$, and the lower panel corresponds to the expected future sensitivity on CR$(\mu-e)_{\rm Al}\leq 10^{-17}$.
Shaded green areas correspond to the stau as the lightest supersymmetric particle ($m_0\ll M_{1/2}$) and no correct electroweak symmetric breaking ($M_{1/2}\ll m_{0}$).
The ATLAS mSUGRA limit covers up to $M_{1/2}=(500\div 800)$ GeV for $m_0\leq 6$ TeV.
We note that the excluded regions, the remaining areas below the corresponding lines, have the typical shape due to LH insertions, see \cite{Lopez-Ibanez:2017xxw}.
Current bounds are already competitive with the experimental limit in Mechanism I, for which the excluded slepton and gluino masses in the less constrained case reach $\widetilde{\ell}\lesssim 800$ GeV and $\widetilde{g} \lesssim 1.5$ TeV\footnote{Hereafter, we omit the cases $z=0$ in Mechanism I and $h_r=0$ in Mechanism II $a$-2 from the discussion since they are already disfavored by the \texttt{Planck} + BAO limit on the total sum of light neutrino masses.}.
These bounds could be increased up to $\widetilde{\ell}\lesssim 3$ TeV and $\widetilde{g} \lesssim 5$ TeV with future sensitivity from $\mu-e$ conversion, (see lower panel in Figure \ref{fig:m0M12_BRmueyExcluded}).
More severe are the constraints imposed by Mechanism II.
In particular, for current bounds, we can infer the following global lower limits: $\widetilde{\ell}\lesssim 1.5\,{\rm TeV}$ and $\widetilde{g}\geq 2.8\,{\rm TeV}$.
They become much stronger considering the future reach of $\mu-e$ conversion which translates into $\widetilde{\ell}\geq 5\,{\rm TeV}$ and $\widetilde{g}\geq 8\,{\rm TeV}$.\\
It means that precision experiments will allow us to put constraints over the supersymmetric spectrum for masses that are beyond the LHC sensitivity in a factor of $3\div 5$.
Conversely, the discovery of SUSY partners in the TeV range will put significant constraints on these simple realizations with LH mediators, which will have to be reformulated in more elaborated scenarios:
for instance, suppressing the LH contributions in the Kähler function or increasing the degrees of freedom in the neutrino sector by considering non-trivial Majorana and Dirac structures simultaneously.

\subsection{Relations among observables}
\label{subsec:resultscombined}
\begin{figure}[t!]
  \centering
  \includegraphics[scale=0.64]{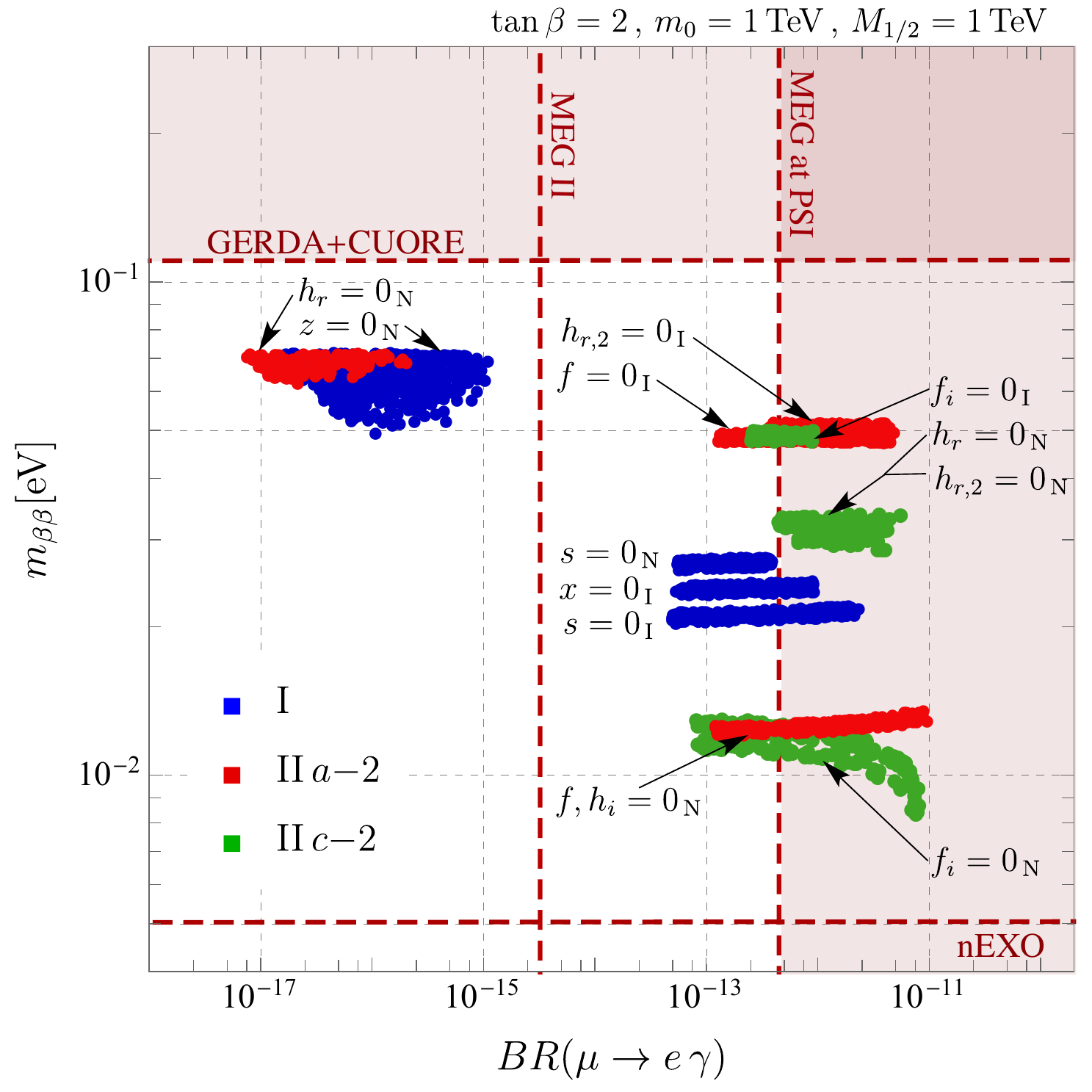}
  \includegraphics[scale=0.64]{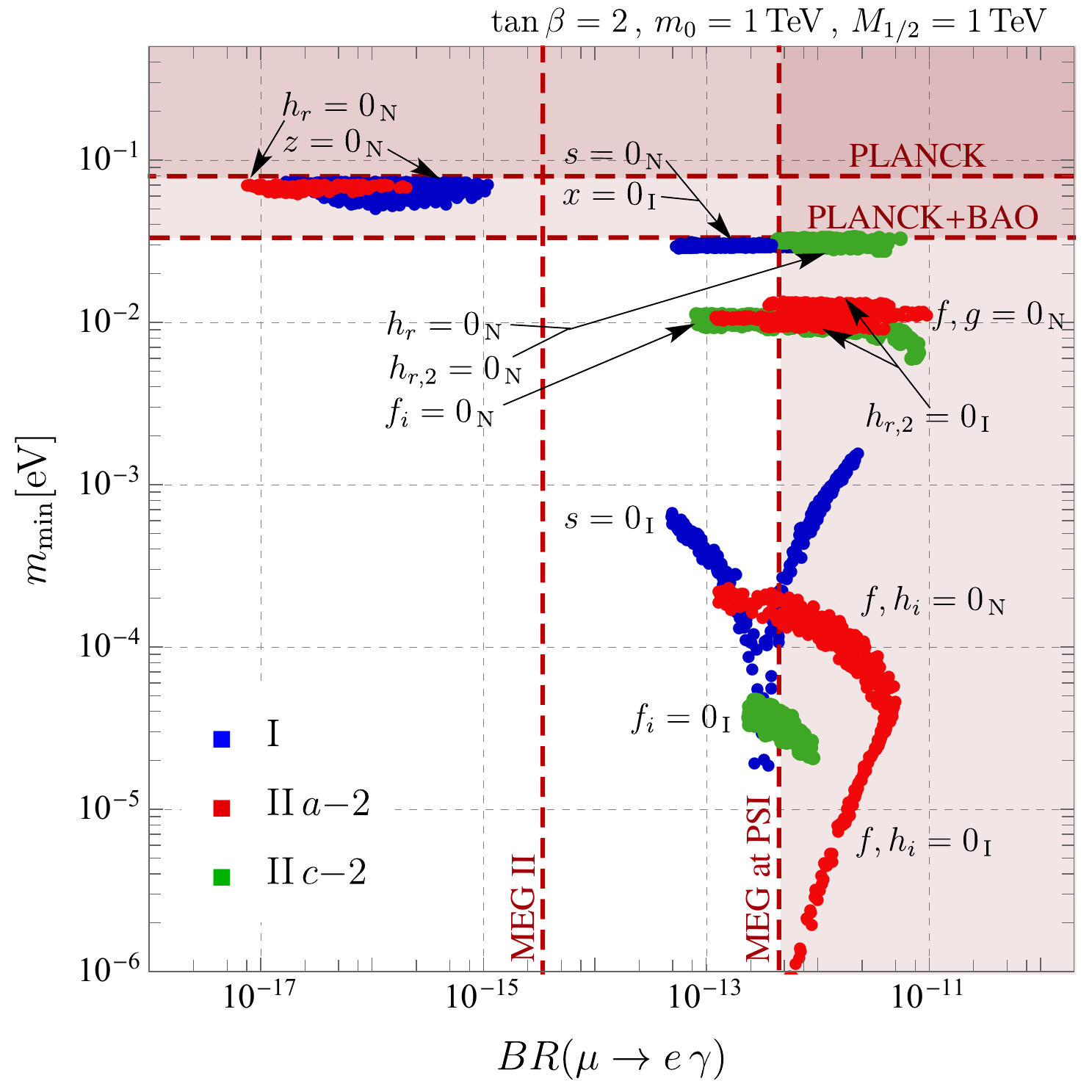}
  \caption{\label{fig:BRmueymbb}\small The quantities $m_{\beta\beta}$ and $m_{\rm min}$ versus the muon flavor violating decay $BR(\mu \rightarrow e \gamma)$. The mSUGRA input parameters are fixed to the conservative values $m_0=M_{1/2}=1\,{\rm TeV},\tan\beta=2$. The shaded regions are already excluded by current bounds.\\ }
\end{figure}
\begin{figure}[h!]
  \centering
  \includegraphics[scale=0.49]{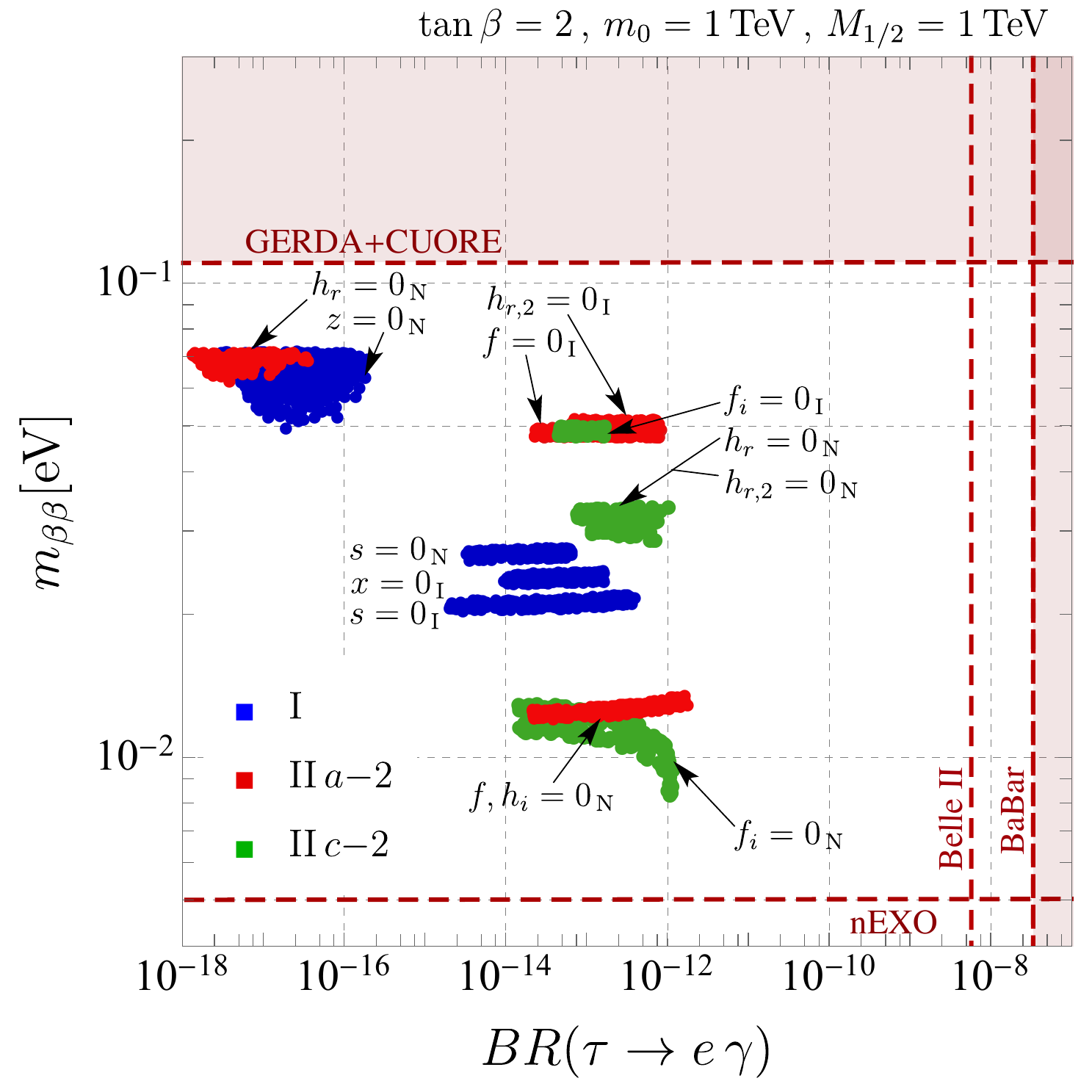}
  \includegraphics[scale=0.49]{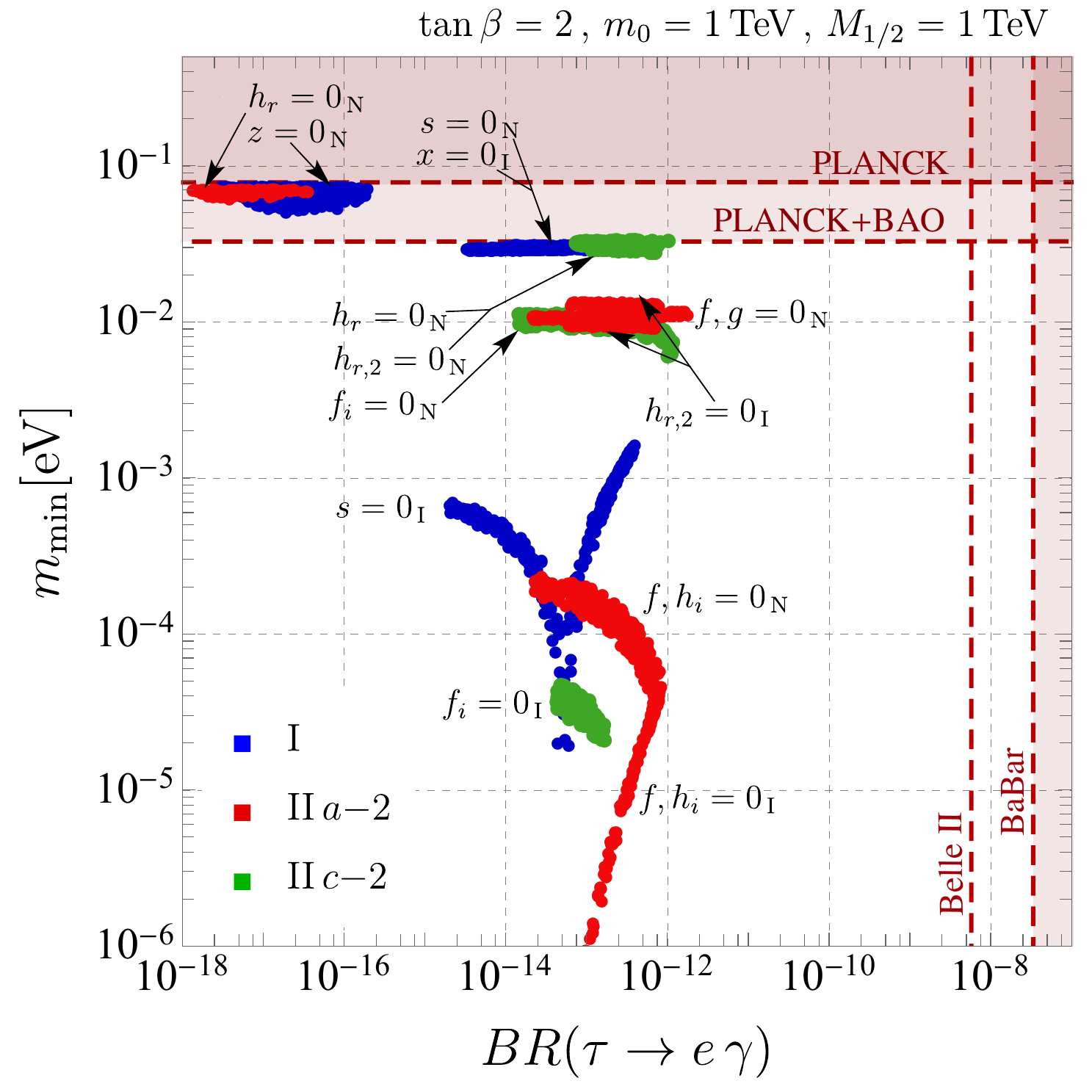}
  \\
  \includegraphics[scale=0.49]{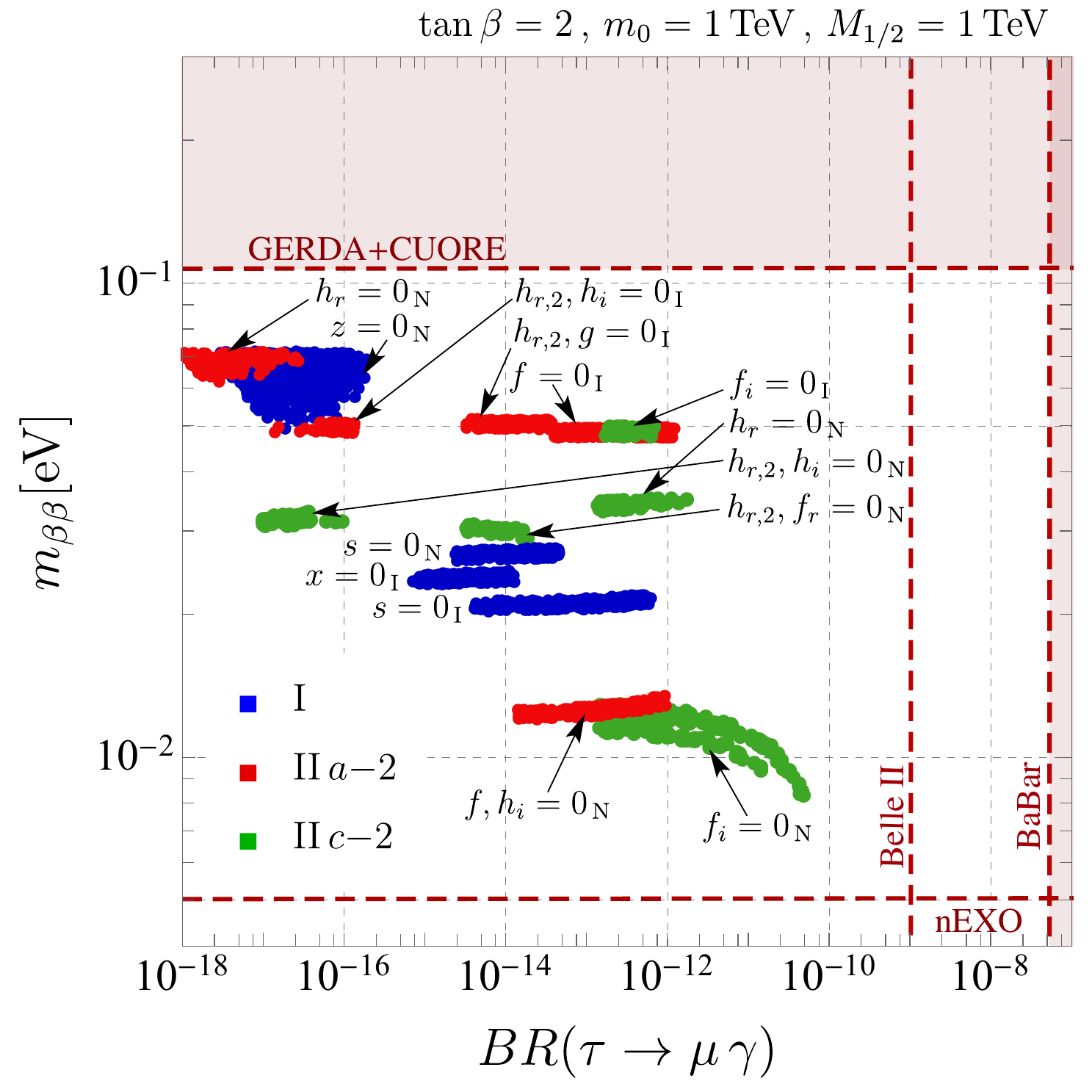}
  \includegraphics[scale=0.49]{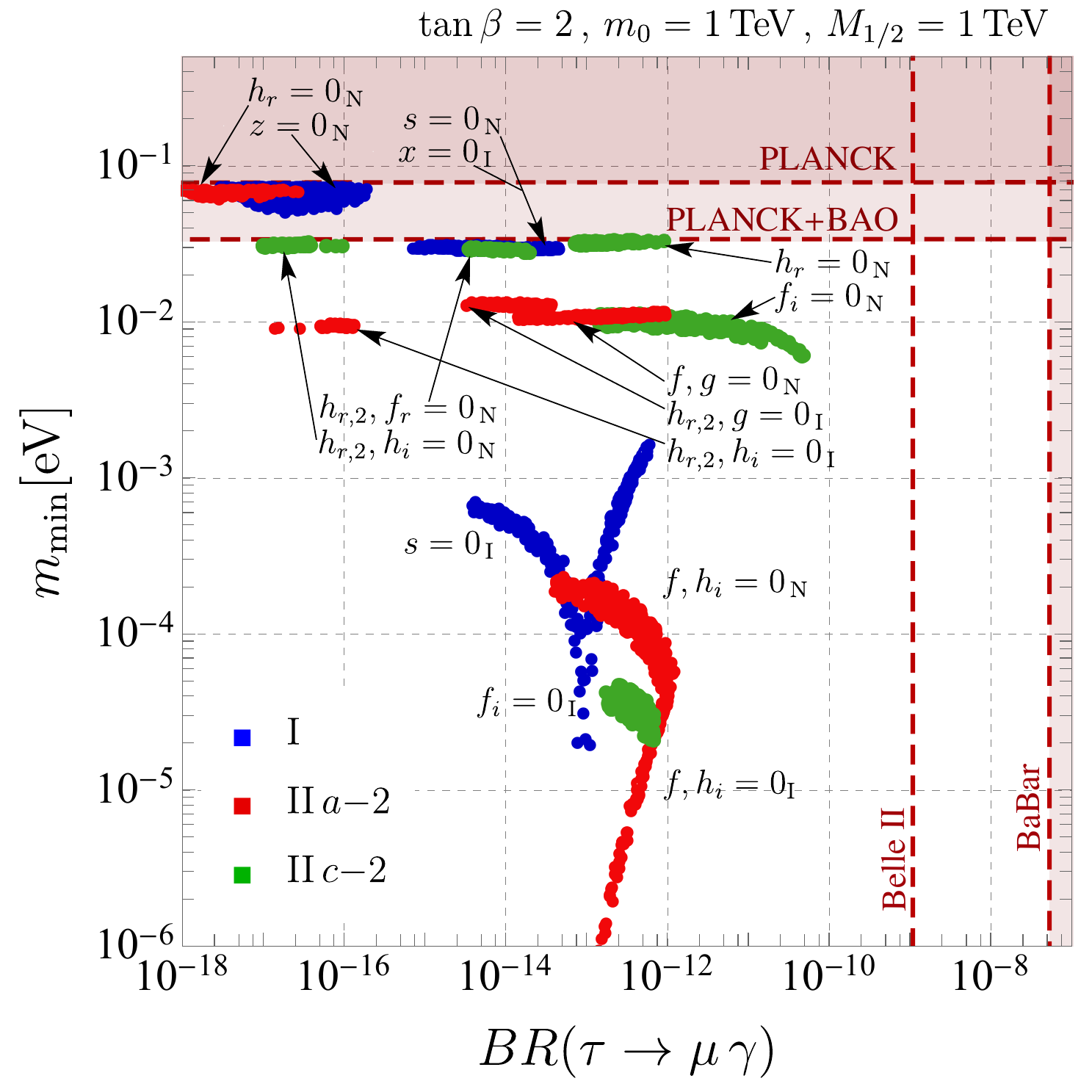}
  \caption{\label{fig:BRtauemuymbb}\small
The quantities $m_{\beta\beta}$ and $m_{\rm min}$ versus the flavor violating decays of the $\tau$: $BR(\tau \rightarrow e \gamma)$ (top panels) and $BR(\tau \rightarrow \mu \gamma)$ (bottom panels).
The mSUGRA input parameters are fixed to the conservative values $m_0=M_{1/2}=1\,{\rm TeV},\tan\beta=2$. The shaded regions are already excluded by current bounds.}
\end{figure}

In this framework it is possible to predict testable relations among the LFV observables in the charged sector, Table~\ref{tab:LFVbounds}, and the neutrino mass observables $m_{\beta\beta}$ and $m_{\rm min}$ discussed in Section \ref{subsec:resultsneutrinos}.
This allows to disentangle cases that are not distinguishable only through the analysis in \cite{DiIura:2015kfa}.
To this end we fix the mSUGRA parameters to $m_0=M_{1/2}=1$ TeV, which would correspond to sleptons and gluinos masses around $1.5$ TeV and $2.5$ TeV respectively, and a very conservative value $\tan\beta=2$, which ensures that the $\tan\beta$-enhanced contributions to the FC processes are minimized.
Then, we study the dependence of both $m_{\beta\beta}$ and $m_{\rm min}$ on ${\rm BR}(\mu \rightarrow e\,\gamma)$ and ${\rm BR}(\tau \rightarrow e\,(\mu)\,\gamma)$.
Our results for each process are shown in Figures \ref{fig:BRmueymbb} and \ref{fig:BRtauemuymbb}.
It is worth emphasizing the interplay of the experimental bounds coming from the two sectors, which acts in a complementary way in constraining the different realizations of the neutrino masses.
In fact, current bounds on LFV observables allow to constrain some regions that otherwise were not testable with present neutrino data, since they predict too low values for $m_{\beta\beta}$ and $m_{\rm min}$ (bottom-right).
In particular, the cases $h_r=0$ and $h_{r,2}=0$ of Mechanism II (both $a$-2 and $c$-2) are severely constrained, while the cases $s=0$ and $x=0$ for Mechanism I, $f=0$ for Mechanism II $a$-2 and $f_i=0$ for Mechanism II $c$-2 result partially excluded.
Moreover, with the expected sensitivity of \texttt{MEG II} \cite{TheMEG:2016wtm} in ${\rm BR} (\mu \to e \gamma)$ all the scenarios detailed before will be completely (dis)proved.
And vice versa, those realizations out of the scope of 
\texttt{MEG II}, namely $z=0$ in Mechanism I and $h_r=0$ in Mechanism II $a$-2, happen to be incompatible with the recent cosmological data presented by the \texttt{Planck} collaboration for the neutrino masses \cite{Aghanim:2018eyx}. 

\section{Conclusions}
\label{sec:conclusions}
In this work we have analyzed the phenomenological consequences of combining $A_5$ and CP as a flavor symmetry in SUSY.
We have focused on the leptonic sector where two residual symmetries, $Z_5$ and $Z_2\times$CP, remain conserved at LO for charged leptons and neutrinos.
In Section \ref{sec:reviewSUSY}, the main effects of introducing a flavor symmetry into SUSY theories have been summarized.
It can been shown that, even in the most conservative scenario, {\it tree-level} FC soft couplings arise when $\Lambda_{\rm SUSY}\gg \Lambda_f$, producing sizable effects in LFV observables.\\
The main features of $A_5$ and CP as a group have been reviewed in Section \ref{sec:reviewA5CP}, where the structure for the lepton mass matrices, the vev of the flavon fields and the leptonic mixing have been derived.
The mass matrix for charged leptons results in being diagonal at LO while, for neutrinos, all possible realizations to generate their masses at tree level have been investigated.
We noticed that two different set of flavons are required to generate the neutrino and charged lepton masses at LO.
Each of them can induce corrections in the opposite sector at NLO, as discussed in Section \ref{subsec:NLOWrongFlavons}.\\
Section \ref{sec:KahlerSoft} has been devoted to compute the minimal set of effective operators entering the Kähler potential and soft masses for LH and RH fields.
Under no additional assumptions about the UV theory, the proposed operators are always present and cannot be avoided through the introduction of additional symmetries.
For the residual symmetries considered here, we have found that the presence of LH mediators is specially relevant and that the flavons associated with neutrino masses also induce flavor violation in the charged-lepton sector.
This allows for a combined analysis of neutrino observables and LFV processes.\\
In Section \ref{subsec:resultsneutrinos}, we have computed our predictions for the neutrino effective mass $m_{\beta\beta}$ versus $m_{\rm min}$.
We observed that the canonical rotation may have some effect over the model parameter $m_{\rm min}$.
Although these variations seem to be difficult to measure, we think that this information could be useful when (re)constructing theoretical models from experimental data.
Regarding charged leptons, in Section \ref{subsec:resultsCL}, we have interpreted our predictions for the LFV processes $\mu \to e \gamma$ and $\mu-e$ conv. in terms of exclusion limits on the plane $\{m_0,M_{1/2}\}$. 
The results depend on the mechanism responsible for the neutrino masses; however,
even in the less restricted realizations, we obtain limits that are competitive with those coming from direct searches of ATLAS on mSUGRA scenarios.
This type of analysis is very useful to indirectly explore the supersymmetric spectrum in concrete models: in the absence of experimental signals of new physics, stringent limits can be set over the superpartner masses for each specific setup; if SUSY is discovered at the TeV scale, the simplified constructions analyzed throughout this article will have difficulties to accommodate it so that more refined scenarios must be considered.
Finally, the common origin of neutrino masses and flavor violation for charged leptons induce testable relations between neutrino and charged-lepton observables. 
This has been expounded in Section \ref{subsec:resultscombined}, where a nice complementarity between both sectors has been found: those realizations difficult to test with neutrino data will be totally probed with the expected sensitivity of \texttt{MEGII} on $\mu \to e \gamma$; conversely, the two scenarios that remain out of the scope of this experiment should be already discarded if the latest limits from \texttt{Planck}+BAO on the total sum of the neutrino masses are imposed.

\section*{Acknowledgments}
OV thanks Università di Roma III for its hospitality. This work has been partially supported under MICINN
Grant FPA2017-84543-P and by the ``Centro de Excelencia Severo Ochoa'' programme under grant SEV-2014-0398. OV acknowledges partial support from the ``Generalitat Valenciana'' grant  PROMETEO2017-033 and AM acknowledges support from La-Caixa-Severo Ochoa scholarship. All Feynman diagrams have been drawn using Jaxodraw \cite{Binosi:2003yf, Binosi:2008ig}.
\appendix
\section{Group Theory}
\label{app:a5xCPgroup}
Here we present the matrix form of the $A_5$ and CP generators, $S_{\bf r}$ and $T_{\bf r}$, for each of the irreducible representations of the group: ${\bf r} =$ \1, \3, $\tr$, \4 and \5 \cite{Feruglio:2011qq}.

\begin{subequations} \label{eqn:genrep}
	\beq \label{eqn:gen1rep} S_{\bf 1} \;=\; e^{i\, \pi} \hspace{1.cm} T_{\bf 1} \;=\; e^{i\, 
		\frac{2\pi}{5}}
    \eeq
	\beq \label{eqn:gen3rep} S_{\bf 3} = \frac{1}{\sqrt{5}}\begin{pmatrix}
		    1      & \sqrt{2}    & \sqrt{2}   \\
		~\sqrt{2}~ & ~-\varphi~ & ~1/\varphi~ \\
		  \sqrt{2} & 1/\varphi    & -\varphi  \\		
		\end{pmatrix}
		\hspace{1.cm} T_{\bf 3} = \begin{pmatrix}
	 	    1 &         0                & 0                         \\
		  ~0~ & ~e^{i\, \frac{2\pi}{5}}~ & 0                         \\
		    0 &         0                & ~e^{i\, \frac{8\pi}{5}}~ \\		
		\end{pmatrix}
    \eeq
	\beq \label{eqn:gen3prep} S_{\bf 3'} = -\frac{1}{\sqrt{5}}\begin{pmatrix}
		    1      & \sqrt{2}    & \sqrt{2}   \\
		~\sqrt{2}~ & ~1/\varphi~ & ~-\varphi~ \\
		  \sqrt{2} & -\varphi    & 1/\varphi  \\		
		\end{pmatrix}
		\hspace{1.cm} T_{\bf 3'} = \begin{pmatrix}
	 	    1 &         0                & 0                         \\
		  ~0~ & ~e^{i\, \frac{4\pi}{5}}~ & 0                         \\
		    0 &         0                & ~e^{-i\, \frac{4\pi}{5}}~ \\		
		\end{pmatrix}
	\eeq
	\beq \label{eqn:gen4rep} S_{\bf 4} = -\frac{1}{5}\begin{pmatrix}
	 	~-\sqrt{5}~ & ~\varphi-3~ & ~\varphi+2~ & ~-\sqrt{5}~ \\
		  \varphi-3 &    \sqrt{5} &    \sqrt{5} &   \varphi+2 \\
		  \varphi+2 &    \sqrt{5} &    \sqrt{5} &   \varphi-3 \\		
		  -\sqrt{5} &   \varphi+2 &   \varphi-3 &   -\sqrt{5} \\
		\end{pmatrix}
		\hspace{0.1cm} T_{\bf 4} = \begin{pmatrix}
		~e^{i\, \frac{2\pi}{5}}~ &         0                &          0                &            0            \\
		           0             & ~e^{i\, \frac{4\pi}{5}}~ &          0                &            0            \\		
		           0             &         0                & ~e^{i\, \frac{6\pi}{5}}~  &            0            \\
		           0             &         0                &          0                & ~e^{i\, \frac{8\pi}{5}}~\\		
		\end{pmatrix}
    \eeq
	\beq \label{eqn:gen5rep} \hspace{-0.35cm} S_{\bf 5} = {\scriptsize \frac{1}{5}\begin{pmatrix}
	 	           ~~~-1 &  \sqrt{6}      & -\sqrt{6}     &  -\sqrt{6}    & -\sqrt{6}    \\
	 	     ~~~\sqrt{6} & ~2-\varphi~    & 2\varphi      & 2(1-\varphi)~ & -(1+\varphi) \\	 	
		     ~-\sqrt{6}~ &  2\varphi      & 1+\varphi     &   2-\varphi   & 2\,(\varphi-1) \\
		       -\sqrt{6} & 2\,(1-\varphi) & 2-\varphi     & 1+\varphi     & -2\varphi    \\		
		       -\sqrt{6} & -(1+\varphi)   & ~2(\varphi-1) & -2\varphi     & 2-\varphi    \\
		\end{pmatrix}}
		\hspace{0.25cm} T_{\bf 5} = {\scriptsize \begin{pmatrix}
		~1~ &            0             &         0                &          0                &            0            \\
		 0  & ~e^{i\, \frac{2\pi}{5}}~ &         0                &          0                &            0            \\
		 0  &            0             & ~e^{i\, \frac{4\pi}{5}}~ &          0                &            0            \\		
		 0  &            0             &         0                & ~e^{i\, \frac{6\pi}{5}}~  &            0            \\
		 0  &            0             &         0                &          0                & ~e^{i\, \frac{8\pi}{5}}~\\		
		\end{pmatrix}}\,.
    \eeq
\end{subequations}

The matrix form of the CP generator, $X_{0,{\bf r}}$, in the irreps ${\bf r}=$ \1, $\tr$, \4 and \5 is:
	\begin{subequations} \label{eqn:X0rep}
	\beq X_{0,{\bf 1}} \;=\; {\bf 1} \hspace{2.cm}
		 X_{0,{\bf 3'}} = \begin{pmatrix}
	 	    								~1~ & ~0~ & ~0~ \\
	 	    							 	 0  &  0  &  1  \\
	 	    								 0  &  1  &  0
	 	    								\end{pmatrix}
	\eeq		
	\beq X_{0,{\bf 4}} = \begin{pmatrix}
	 					~0~ & ~0~ & ~0~ & ~1~ \\
	 					 0  & ~0~ & ~1~ & ~0~ \\
	 					 0  & ~1~ & ~0~ & ~0~ \\
	 					 1  & ~0~ & ~0~ & ~0~
	 					\end{pmatrix} \hspace{1.cm}
	 	 X_{0,{\bf 5}} = \begin{pmatrix}
	 					~0~ & ~0~ & ~0~ & ~0~ & ~1~ \\
	 					~0~ &  0  & ~0~ & ~1~ & ~0~ \\
	 					~0~ &  0  & ~1~ & ~0~ & ~0~ \\
	 					~0~ &  1  & ~0~ & ~0~ & ~0~ \\
	 					~1~ &  0  & ~0~ & ~0~ & ~0~
	 					\end{pmatrix}\,.
	\eeq
\end{subequations}
For ${\bf r}=$ \3, $X_{0,{\bf 3}}$ has been specified in Eq.\eqref{eqn:P23}.

\section{\boldmath $U_{\rm PMNS}$ parametrization}
\label{app:upmns}
We use the following convention for the PMNS matrix:
\begin{align} \label{PMNS_definition_general}
 U_{\rm PMNS} = \tilde{U} \ {\rm diag}\{1, e^{i \alpha/2}, e^{i (\beta/2 + \delta)}\},
\end{align}
where $\alpha$ and $\beta$ are the Majorana phases and $\tilde{U}$ the CKM-like parametrization of the mixing matrix,
\begin{align}
	\tilde{U} = \begin{pmatrix}
         1  &     0    &    0     \\
        ~0~ & \quad c_{23}~ & ~s_{23}~ \\
         0  &  -s_{23} &  c_{23}  \\
	\end{pmatrix}
	\begin{pmatrix}
        c_{13} & 0 & ~s_{13}\, e^{-i \delta} \\
           0   & 1 & 0 \\
        -s_{13}\, e^{i \delta}~ & 0 & c_{13} \\
	\end{pmatrix}
	\begin{pmatrix}
        \quad c_{12}~  & ~s_{12}~ & ~0~ \\
        -s_{12} & c_{12} & 0 \\
        0 &  0 & 1 \\
	\end{pmatrix}
\end{align}
with $c_{ij} \equiv \cos\theta_{ij}$, $s_{ij} \equiv \sin\theta_{ij}$ and $\delta$ the Dirac CP phase.
All the angles are in the first quadrant $\theta_{ij} \in [0, \pi/2]$ and can be extracted using the PMNS matrix elements as:
 \begin{align}
 \label{mixing_angles_from_PMNS}
  \sin^2\theta_{12} = \frac{|U_{12}|^2}{1-|U_{13}|^2}, \qquad \sin^2\theta_{13} = |U_{13}|^2, \qquad \sin^2\theta_{23} = \frac{|U_{23}|^2}{1-|U_{13}|^2}\,.
 \end{align}
The Majorana phases can be derived from the numerical PMNS mixing matrix taking into account the unphysical phases described by the diagonal matrix that multiplies the $U_{\rm PMNS}$ from the left, ${\rm Diag}\{e^{i \delta_e}, e^{i \delta_\mu}, e^{i \delta_\tau}\}$.
Those can be eliminated with a redefinition of the charged lepton fields.
We can obtain the Majorana phases as
\begin{align}
 \alpha &= 2 \arg \left\{\frac{U_{12}}{U_{11}} \right\} \qquad\qquad\qquad \beta = 2 \arg \left\{\frac{U_{13}}{U_{11}}\right\}.
\end{align}
For sake of completeness we report the values of the unphysical phases
\begin{align}
 \delta_e = \arg\{U_{11}\}\quad 
 \delta_\mu = \arg \left\{U_{23}e^{-i (\beta/2 + \delta) } \right\}\quad
 \delta_\tau =\arg \left\{U_{33}e^{-i (\beta/2 + \delta) }\right\}.
\end{align}
A similar procedure is discussed in \cite{Antusch:2003kp} using a slightly different parametrization for the PMNS matrix.

\newpage
\section{Flavor Observables}
\label{app:observables}
Here we detail the numerical values for the neutrino and charged-lepton observables considered in the analysis.
Table \ref{tab:neutrinoBounds} collects the latest results for the global fit of the neutrino oscillation parameters performed by the NuFIT Collaboration in \cite{Esteban:2018azc}.
In Table \ref{tab:Cosmobounds}, we show the most stringent bounds over the neutrino effective masses, $m_{\beta\beta}$ and $m_\beta$, and the total sum of the neutrino masses (first column) and the future sensitivity on those observables (second column).
Table \ref{tab:LFVbounds} refers to current and expected future bounds over the flavor-changing processes: $l_i \to l_j \gamma$, $l_i \to 3l_j$ and $\mu-e$ conversion.

\begin{table}[h!]
\centering 
{\small
\renewcommand{\arraystretch}{1.4}%
\resizebox{\columnwidth}{!}{
\begin{tabular}{|c|c|c|}
\hline
{\bf Parameter} & {\bf NH $\boldsymbol{3\sigma}$-Range}    & {\bf IH $\boldsymbol{3\sigma}$-Range}\\
\hline 
\hline 
$\sin^2\theta_{12}$ & $\hspace{10mm}(2.75\div 3.50)\times 10^{-1}\hspace{10mm}$& $\hspace{10mm}(2.75\div 3.50)\times 10^{-1}\hspace{10mm}$ \\
$\sin^2\theta_{23}$ & $(4.28\div 6.24)\times 10^{-1}$ & $(4.33\div 6.23)\times 10^{-1}$ \\
$\sin^2\theta_{13}$ & $(2.044\div 2.437)\times 10^{-2}$ & $(2.067\div 2.462)\times 10^{-2}$ \\
$\Delta m^2_{21}$ [eV$^2$] & $(6.79\div 8.01)\times 10^{-5}$  & $(6.79\div 8.01)\times 10^{-5}$ \\
$\Delta m^2_{3j}$ [eV$^2$] & $(2.431\div 2.622)\times 10^{-3}$ & $ -(2.416\div 2.606)\times 10^{-3}$ \\
$r_j$ & $(2.590 \div 3.295)\times 10^{-2}$   & $-(2.606\div 3.320)\times 10^{-2}$ \\
\hline
\end{tabular}}}
\captionsetup{width=0.98\textwidth}
\caption{\label{tab:neutrinoBounds} \s
Latest results for the global fit of the neutrino oscillation data from NuFIT 4.0 (2018), \href{http://www.nu-fit.org}{\texttt{http://www.nu-fit.org}} \cite{Esteban:2018azc} considered in the analysis.
For NO, $j=1$, while for IO, $j=2$.}
\end{table}

\begin{table}[h!]
\centering 
{\small
\renewcommand{\arraystretch}{1.4}%
\resizebox{\columnwidth}{!}{
\begin{tabular}{|c|c|c|}
\hline
{\bf Observable} & {\bf Current Bound}    & {\bf Future Bound}     \\
\hline 
\hline 
$m_{\beta\beta}$\,[eV]   &   0.11  {\s @\,90\%}  (\texttt{CUORE}\cite{Alduino:2017ehq}) &  0.005 {\s @\,90\%} (\texttt{nEXO}\cite{Albert:2017hjq})\\
$m_\beta$\,[eV] & $-$  & 0.02 {\s @\,90\%} (\texttt{KATRIN}\cite{Angrik:2005ep, SejersenRiis:2011sj}) \\
\multirow{2}{*}{$\sum m_j$\,[eV]} & 0.26 {\s @\,95\%} (\texttt{Planck}\cite{Aghanim:2018eyx}) & \multirow{2}{*}{0.062 {\s @\,68\%} (\texttt{CORE}+BAO\cite{DiValentino:2016foa})}\\
& 0.12 {\s @\,95\%} (\texttt{Planck}+BAO\cite{Aghanim:2018eyx}) & \\
\hline
\end{tabular}}}
\captionsetup{width=0.98\textwidth}
\caption{\label{tab:Cosmobounds} \s
Current and future bounds on the neutrino observables: $m_{\beta\beta}$, $m_\beta$ and the total sum of neutrino masses.
Note that the latest limits from \texttt{Planck} fall in the quasi-degenerate regime for the neutrino spectrum.
In that case, these bounds can be translated into bounds over the lightest mass eigenstate as: $m_{\min} \lesssim 0.09$ eV for \texttt{Planck} data and $m_{\rm min} \lesssim 0.04$ eV for \texttt{Planck}+BAO. }
\end{table}

\begin{table}[t!]
{\small
\centering 
\renewcommand{\arraystretch}{1.4}%
\resizebox{\columnwidth}{!}{
\begin{tabular}{|c|c|c|}
\hline
{\bf LFV process} & {\bf Current Bound}    & {\bf Future Bound}     \\
\hline 
\hline 
BR$(\mu  \to e \gamma)$  & $4.2 \times 10^{-13}$ (\texttt{MEG at PSI}\cite{TheMEG:2016wtm}) & $6 \times 10^{-14}$ \texttt{(MEG\,II} \cite{Baldini:2018nnn}) \\
BR$(\mu  \to e e e)$        & $1.0 \times 10^{-12}$ (\texttt{SINDRUM}\cite{Bellgardt:1987du}) & ~~\quad $10^{-16}$ (\texttt{Mu3e}\cite{Blondel:2013ia}) \\
CR$(\mu-e\,)_{A_l}$        & $-$ & ~~\quad $10^{-17}$ (\texttt{Mu2e}\cite{Bartoszek:2014mya}, \texttt{COMET}\cite{Blondel:2013ia}) \\
BR$(\tau \to e \gamma)$     & $3.3 \times 10^{-8}$ (\texttt{BaBar}\cite{Aubert:2009ag})  & \qquad $5\times 10^{-9}$ (\texttt{Belle\,II}\cite{Aushev:2010bq}) \\
BR$(\tau \to \mu \gamma)$   & $4.4 \times 10^{-8}$  (\texttt{BaBar}\cite{Aubert:2009ag}) & \qquad $10^{-9}$ (\texttt{Belle\,II}\cite{Aushev:2010bq}) \\
BR$(\tau \to e e e)$        & $2.7 \times 10^{-8}$  (\texttt{Belle}\cite{Miyazaki:2011xe}) & \qquad $5\times10^{-10}$ (\texttt{Belle\,II}\cite{Aushev:2010bq}) \\
BR$(\tau \to \mu \mu \mu)$  & $2.1 \times 10^{-8}$ (\texttt{Belle}\cite{Miyazaki:2011xe})  & \qquad $5\times10^{-10}$ (\texttt{Belle\,II}\cite{Aushev:2010bq}) \\
\hline
\end{tabular}}}
\captionsetup{width=0.98\textwidth}
\caption{\label{tab:LFVbounds} Relevant LFV processes considered in our analysis. 
}
\end{table}

\newpage
\section{Bounds on Mass Insertions}
\label{app:deltabounds}
Table \ref{tab:deltasRanges} collects the obtained intervals for the LL mass insertions after imposing all the constraints regarding lepton mixing and masses.
Specifically, we report the off-diagonal elements of the soft mass matrix for LH sleptons divided by the mSUGRA parameter $m_0^2$, {\it i.e.} the value of the MI at the scale of flavor breaking, before RGE evolution.
The allowed ranges are specified for each realization of the neutrino mass mechanisms analyzed in this work.
In Section \ref{subsec:resultsCL}, the minimum of those intervals has been employed to constrain the mSUGRA plane $m_0-M_{1/2}$ for a representative value of $\tan\beta=10$.

{\tablinesep=7pt
\begin{table}[h!]
\centering
\renewcommand{\arraystretch}{1.}%
\resizebox{\columnwidth}{!}{
\begin{tabular}{|c|c|c|c c c| }
\hline
\multicolumn{2}{|c}{\bf\hspace{6mm} Mechanism} &  & {\boldmath $\delta^{\rm}_{12}$} & {\boldmath $\delta^{\rm}_{13}$} & {\boldmath $\delta^{\rm}_{23}$} \\[1pt]
\hline
\hline
\multirow{ 4}{*}{I} & $z=0$ & NH & {\s $(3.9 \div 32)\times10^{-5}$} & {\s $(3.9 \div 32)\times10^{-5}$} & {\s $(3.9 \div 32)\times10^{-5}$} \\[1pt]
\cline{2-6}
& $x=0$ & IH & {\s $(2.2\div 9.2)\times10^{-3}$} & {\s $(2.2\div 9.2)\times10^{-3}$} & {\s $(6\div 26)\times10^{-4}$} \\[1pt]
\cline{2-6}
&\multirow{ 2}{*}{$s=0$} & NH & {\s $(1.3\div 5.8)\times10^{-3}$} & {\s $(1.3\div 5.8)\times10^{-3}$} & {\s $(1.1\div 4.8)\times10^{-3}$} \\
&	 & IH & {\s $(1\div 15)\times10^{-3}$} & {\s $(1\div 14)\times10^{-3}$} & {\s $(1.5\div 18)\times10^{-3}$} \\[1pt]
\hline
\hline
\multirow{ 6}{*}{II $a$-2} & $ f,g=0$ & IH & {\s $(3.8\div 21)\times 10^{-3}$} & {\s $(3.8\div 21)\times 10^{-3}$} & {\s $(4.6\div 23)\times 10^{-3}$} \\[5pt]
\cline{2-6}
&\multirow{ 2}{*}{$f,h_i=0$} & NH & {\s $(3.3\div 29)\times 10^{-3}$} & {\s $(3.3\div 30)\times 10^{-3}$} & {\s $(2.7\div 22)\times 10^{-3}$} \\
&  & IH & {\s $(3.4\div 21)\times 10^{-3}$} & {\s $(3.4\div 20)\times 10^{-3}$} & {\s $(4.5\div 25)\times 10^{-3}$} \\[1pt]
\cline{2-6}
& $ h_r,g=0$ & NH & {\s $(2.6\div 14)\times 10^{-5}$} &  {\s $(2.6\div 14)\times 10^{-5}$} &  {\s $(2.3\div 12)\times 10^{-5}$} \\[1pt]
\cline{2-6}
& $ h_{r,2},g=0$ & IH & {\s $(5.8\div 20)\times 10^{-3}$} & {\s $(5.8\div 20)\times 10^{-3}$} & {\s $(1.3\div 4.3)\times 10^{-3}$} \\[1pt]
\cline{2-6}
& $h_{r,2},h_i=0$ & IH & {\s $(5.6\div 19)\times 10^{-3}$} & {\s $(5.6\div 19)\times 10^{-3}$} & {\s $(8.5\div 26)\times 10^{-5}$} \\[1pt]
\hline
\hline
\multirow{ 8}{*}{II $c$-2} &\multirow{2}{*}{$f_i,h_i=0$} & NH & {\s $(2.7\div 27)\times 10^{-3}$} & {\s $(2.7\div 24)\times 10^{-3}$} & {\s $(0.9\div 16)\times 10^{-2}$} \\
& & IH & {\s $(4.8\div 9.1)\times 10^{-3}$} & {\s $(4.8\div 9.2)\times 10^{-3}$} & {\s $(9.4\div 19)\times 10^{-3}$} \\[1pt]
\cline{2-6}
&\multirow{ 2}{*}{$f_i,f_r=0$}& NH & {\s $(2.8\div 22)\times 10^{-3}$} & {\s $(2.8\div 21)\times 10^{-3}$} & {\s $(8.6\div 86)\times 10^{-3}$} \\
&  & IH & {\s $(4.7\div 8)\times 10^{-3}$} & {\s $(4.8\div 8.1)\times 10^{-3}$} & {\s $(1.1\div 1.9)\times 10^{-2}$} \\[1pt]
\cline{2-6}
& $h_r,h_i=0$ & NH & {\s $(6.1\div 15)\times 10^{-3}$} & {\s $(6.1\div 15)\times 10^{-3}$} & {\s $(6\div 15)\times 10^{-3}$} \\[1pt]
\cline{2-6}
& $h_r,f_r=0$ & NH & {\s $(6.3\div 22)\times 10^{-3}$} & {\s $(6.3\div 23)\times 10^{-3}$} & {\s $(6.2\div 22)\times 10^{-3}$} \\[1pt]
\cline{2-6}
& $h_{r,2},h_i=0$ & NH & {\s $(8.1\div 20)\times 10^{-3}$} & {\s $(8.1\div 20)\times 10^{-3}$} & {\s $(0.8\div 21)\times 10^{-6}$} \\[1pt]
\cline{2-6}
& $h_{r,2},f_r=0$ & NH & {\s $(6.3\div 22)\times 10^{-3}$} & {\s $(6.3\div 23)\times 10^{-3}$} & {\s $(6.2\div 22)\times 10^{-3}$} \\[1pt]
\hline
\end{tabular}}
\caption{\label{tab:deltasRanges}
Obtained intervals for the LL mass insertions: $\delta_{12},\, \delta_{13},\, \delta_{23}$.
The allowed ranges are specified for each realization of the neutrino mass mechanisms analyzed in this work.}
\end{table}}

\newpage
\bibliographystyle{JHEP}
\bibliography{references}
\end{document}